\documentclass[twocolumn]{aastex7}

\usepackage{comment}

\newcommand{\code}[1]	{\textsf{\small #1}}%		# for in main text
\newcommand{\ha}{\hbox{H$\alpha$}}
\newcommand{\hb}{\hbox{H$\beta$}}

\newcommand{\nii}{\hbox{[N\,{\sc ii}]}}
\newcommand{\oiii}{\hbox{[O\,{\sc iii}]}}

\defcitealias{Tempel14}{T14}
\defcitealias{Banerji10}{B10}

\received{December 5, 2024}
\revised{March 18, 2025}
\accepted{April 25, 2025}
\submitjournal{ApJ}
\begin{document}

\title{Morphological Feature Distances among the Spectral Types of SDSS Galaxies}

\correspondingauthor{Duho Kim}

\author[orcid=0000-0001-5120-0158, gname=Duho, sname=Kim]{Duho Kim}
\affiliation{Astronomy and Space Science Department, Chungnam National University, Daehak-ro 99, Yuseong-gu Daejeon 34134, Republic of Korea}
\email[show]{duhokim81@gmail.com}

\author[orcid=0000-0003-2939-8668, gname=Garreth, sname=Martin]{Garreth Martin}
\affiliation{School of Physics and Astronomy, University of Nottingham, University Park, Nottingham NG7 2RD, UK}
\email{garreth.martin@nottingham.ac.uk}

%% Use the \collaboration command to identify collaborations. This command
%% takes an optional argument that is either a number or the word "all"
%% which tells the compiler how many of the authors above the command to
%% show. For example "\collaboration[all]{(DELVE Collaboration)}" wil include
%% all the authors above this command.
%%
%% Mark off the abstract in the ``abstract'' environment. 
\begin{abstract}

This study investigates the morphological feature distances among various spectral types of galaxies from the Sloan Digital Sky Survey, including strong and weak active galactic nuclei (AGN), quasi-stellar objects (QSO), quiescent, and star-forming galaxies. 
We evaluated the clustering and relative distances of these spectral types in the multidimensional morphological feature space. The results indicate that AGN and QSOs are more closely associated with quiescent galaxies than with star-forming ones, indicating a potential regulation of star formation by AGN activity. Furthermore, the analysis underlines the role of AGN feedback in the dwarf regime having $\sim$10--50\% closer distances from AGN types to the quiescent type than to the star-forming type in the dwarf regime $-18 > M_r > -20$, compared to $<$15\% closer in the massive regime $M_r < -21$.
The continuous probability analysis of spectral types being Hubble types upholds the distance analysis results having a range of the probability distribution of AGN types similar to the quiescent type, especially in dwarf galaxies.

\end{abstract}

%% Keywords should appear after the \end{abstract} command. 
%% The AAS Journals now uses Unified Astronomy Thesaurus (UAT) concepts:
%% https://astrothesaurus.org
%% You will be asked to selected these concepts during the submission process
%% but this old "keyword" functionality is maintained in case authors want
%% to include these concepts in their preprints.
%%
%% You can use the \uat command to link your UAT concepts back its source.

%% From the front matter, we move on to the body of the paper.
%% Sections are demarcated by \section and \subsection, respectively.
%% Observe the use of the LaTeX \label
%% command after the \subsection to give a symbolic KEY to the
%% subsection for cross-referencing in a \ref command.
%% You can use LaTeX's \ref and \label commands to keep track of
%% cross-references to sections, equations, tables, and figures.
%% That way, if you change the order of any elements, LaTeX will
%% automatically renumber them.

\section{Introduction} \label{sec:intro}

Compared to the dark matter halo mass function predicted by the $\Lambda$CDM cosmology, the galaxy stellar mass function from large surveys such as the Sloan Digital Sky Survey \citep[SDSS;][]{Gunn98,York00} exhibits a greater abundance of galaxies at stellar masses around that of the Milky Way in the nearby Universe \citep{Guo10}. In other words, galaxies residing within dark matter halos of mass similar to that of the Milky Way ($\sim$10$^{12} \, M_{\sun}$) have been forming stars most efficiently across cosmic time, and galaxies in dark matter halos with masses more or less massive than this have been hindered from forming stars. 

Stellar and active galactic nucleus (AGN) feedback (FB) has been introduced and implemented in hydrodynamical simulations \citep{Springel05_2, Li24, Ni24, Hazenfratz25} and semianalytic models as the hindering process of star formation, with stellar and AGN FB being the dominant processes affecting star formation in dark matter halos less and more massive than $\sim$10$^{12} \, M_{\sun}$, respectively \citep{Springel05_1, Bower06, Crain&Voort23}. In a shallower gravitational potential, FB processes from stars through ionizing radiation, stellar winds, supernova explosions, and outflows from young stars can be enough to blow the interstellar medium away \citep{Dekel&Silk86, Martin25}. In massive halos, the accretion disk around supermassive black holes (SMBHs) more effectively heats and expels gas, preventing it from cooling and forming stars \citep{Silk&Rees98, Fabian12, Dubois16, Kaviraj17}. A tight correlation between the masses of SMBHs and the velocity dispersion of stars in host galaxies is thought to be the outcome of SMBH-regulated star formation history of host galaxies \citep{Gebhardt00, Zhang25}. Processes such as tidal stripping and ram pressure stripping in dense environments (e.g., galaxy clusters) can also remove gas from galaxies, affecting their ability to form stars. However, the relative importance of each mode through cosmic time remains an open question \citep{Nyland18}.

Evidence of FB processes has been observed in the properties of their host galaxies. \citet{Kauffmann03} found that the host galaxies of the low-\oiii-luminosity AGN have stellar populations similar to normal early-type galaxies and younger mean stellar ages in high-\oiii-luminosity AGN at 0.02\,$<$\,$z$\,$<$\,0.3. 
%the highest fraction of luminous  AGN host galaxies in the ``green valley'', $0.5 < U-V < 1.0$, with S\'ersic index values $n$\,$>$\,2.5 at 0.4\,$\leq$\,$z$\,$\leq$\,1.1.
AGN host galaxies have visible colors near the green valley, and this can be interpreted as a transition from a blue cloud to a red sequence by star formation quenching \citep{Schawinski09, Povic12} or dust-reddened star formation \citep{Cardamone10}. \citet{Gabor09} found a broad distribution of X-ray-selected AGN host galaxies' morphologies to be intermediate between the bulge- and disk-dominated regimes without any indication of major mergers \citep[see][]{Kimbrell24}.
%Moreover, the study by \citet{Reviglio06} pointed out the connection between emission-line radio AGNs and late-type galaxies, while absorption-line radio AGNs are associated with early-type morphology. 
\citet{Silverman08} found that the color evolution of AGN host galaxies with moderate X-ray luminosity (41.9\,$\leq$\,$\log L_{0.5-8.0\,keV}$\,$\leq$\,43.7, $M_V\,\lesssim\,-21$) goes from blue (rest-frame $U-V < 0.7$) at $z$\,$\gtrsim$\,0.8 to the majority of host galaxies being along the red sequence at $z$\,$\lesssim$\,0.6 and that an enhancement of AGN activity takes place in large-scale structures.

The large volume of available observational data, combined with the complex interplay of various processes shaping galaxy properties, presents a significant challenge in establishing a direct relationship between galaxy properties and FB processes, which is additionally complicated by the indirect nature of the evidence supporting the importance of feedback. Technological advances in computing power and algorithms can facilitate dealing with a large volume of data. In particular, unsupervised machine learning has been shown to be a powerful tool for analyzing the increasing size of feature parameter space through dimensional reduction, namely, Principal Component Analysis \citep[\code{PCA};][]{Hotelling36}, $k$-means \citep{MacQueen67}, self-organizing maps \citep{Kohonen82} and t-distributed Stochastic Neighbor Embedding \citep[\code{t-SNE};][]{vanderMaaten08}, and so on \citep{Hocking17, Martin20}. 

In this study, we investigate the evolution of the distribution of spectrally classified SDSS galaxies, including strong\,AGN, weak\,AGN, quasi-stellar object (QSO), quiescent (Q), and star-forming (SF), in the morphological feature space by calculating the distances among the spectral types and clustering them on the \code{t-SNE} maps. 
%As Section~\ref{sec:assign} details, the spectral classification uses spectral-line measurements only which are independent of the photometric feature space. If the AGNs are primarily responsible for quenching star formation in massive galaxies as semi-analytic models are currently assuming, galaxies hosting AGN would have transitionaling photometric features from star-forming galaxies to red and quiescent galaxies.

The rest of the article is divided into five sections as follows. In Section~\ref{sec:data}, we present an overview of the catalog data that we use in this study. In Section~\ref{sec:assign}, we explain how we classified the spectral types of SDSS galaxies. In Section~\ref{sec:clustering}, we describe how we perform multidimensional analysis and its results. In Section~\ref{sec:dis}, we contextualize the results within the broader AGN FB framework. In Section~\ref{sec:summary}, we summarize our findings in this study.

\section{The Catalog Data} \label{sec:data}

We used SDSS galaxy catalogs from \citet[hereafter \citetalias{Tempel14}]{Tempel14}. They provide seven volume-limited samples with different magnitude ($M_r$) cuts: $-$18.0, $-$18.5, $-$19.0, $-$19.5, $-$20.0, $-$20.5, and $-$21.0 (see their Figure~2 and Table~1). Larger (dimmer) magnitude cuts result in a lower (closer) redshift limit because of the flux limits of the observations. Each galaxy can appear in multiple volume-limited samples from \citetalias{Tempel14} because each sample contains all galaxies brighter than its magnitude cut. For example, a galaxy with $M_r$\,$=$\,$-$22 and $z$\,$=$\,0.01 is included in all samples. To avoid redundancy, we divide the volume-limited sample into bins of absolute $r$-band magnitude and redshift (see Table~\ref{tab:sample}). 

We cross-matched the catalog \citetalias{Tempel14} with the \code{PhotoObj}, \code{SpecObj}, and \code{galSpecLine} data tables obtained from the SDSS Query site.\footnote{\url{https://skyserver.sdss.org/CasJobs/}} We selected the \code{Context} column Data Release 18 \citep{sdssdr18} and included objects with \code{SpecObj.class} value \code{GALAXY} or \code{QSO}, and \code{SpecObj.zWarning}\,$=$\,0.

\begin{deluxetable*}{lcccccccc}[!ht]
\tablecaption{SDSS galaxy samples with absolute magnitude and redshift limits\label{tab:sample}}
\tablewidth{0pt}
\tablehead{
\colhead{Sample} & \colhead{$M_{r,\mathrm{lim}}$$^a$} & \colhead{$z_{\mathrm{lim}}$$^b$} &  \colhead{$N_{\mathrm{Q}}$$^c$} & \colhead{$N_{\mathrm{SF}}$$^d$} & \colhead{$N_{\mathrm{weak\,AGN}}$$^e$} & \colhead{$N_{\mathrm{strong\,AGN}}$$^f$} & \colhead{$N_{\mathrm{QSO}}$$^g$} &\colhead{$N_{\mathrm{total}}$$^h$} 
}
\startdata
gs180       & $-$18\,$>$\,$M_r$\,$>$\,$-$18.5  & 0.000--0.045  & 2405 (59)     & 8471 (59)     & 54    & 1     & 4     & 10935 (177) \\
gs185       & $-$18.5\,$>$\,$M_r$\,$>$\,$-$19  & 0.000--0.057  & 4784 (206)    & 13260 (206)   & 178   & 9     & 19    & 18250 (618) \\
gs190-05    & $-$19\,$>$\,$M_r$\,$>$\,$-$19.5  & 0.000--0.050  & 3761 (255)    & 7519 (255)    & 220   & 14    & 21    & 11535 (765)  \\
gs190-07    & $-$19\,$>$\,$M_r$\,$>$\,$-$19.5  & 0.050--0.071  & 6763 (273)    & 12952 (273)   & 217   & 25    & 31    & 19988 (819) \\
gs195-05    & $-$19.5\,$>$\,$M_r$\,$>$\,$-$20  & 0.000--0.050  & 3866 (350)    & 5488 (350)    & 311   & 14    & 25    & 9704 (1050) \\
gs195-09    & $-$19.5\,$>$\,$M_r$\,$>$\,$-$20 & 0.050--0.089  & 20415 (974)   & 24330 (974)   & 731   & 127   & 116   & 45719 (2922) \\
gs200-05    & $-$20\,$>$\,$M_r$\,$>$\,$-$20.5  & 0.000--0.050  & 3295 (445)    & 3643 (445)    & 370   & 36    & 39    & 7383 (1335) \\
gs200-11    & $-$20\,$>$\,$M_r$\,$>$\,$-$20.5  & 0.050--0.110  & 36319 (2000)  & 29161 (2000)  & 1302  & 440   & 357   & 67579 (6099) \\
gs205-05    & $-$20.5\,$>$\,$M_r$\,$>$\,$-$21  & 0.000--0.050  & 2254 (362)    & 2110 (362)    & 297   & 33    & 32    & 4726 (1086) \\
gs205-10    & $-$20.5\,$>$\,$M_r$\,$>$\,$-$21  & 0.050--0.100  & 19478 (1764)  & 12159 (1764)  & 1116  & 345   & 303   & 33401 (5292) \\
gs205-14    & $-$20.5\,$>$\,$M_r$\,$>$\,$-$21 & 0.100--0.136  & 36708 (1760)  & 16742 (1760)  & 730   & 646   & 384   & 55210 (5280) \\
gs210-05    & $-$21\,$>$\,$M_r$        & 0.000--0.050  & 1485  (326)   & 1170 (326)    & 267   & 30    & 29    & 2981 (978) \\
gs210-10    & $-$21\,$>$\,$M_r$        & 0.050--0.100  & 14561 (1754)  & 6461 (1754)   & 1151  & 323   & 280  & 22776 (5262) \\
gs210-17    & $-$21\,$>$\,$M_r$       & 0.100--0.168  & 71281 (2000)  & 17958 (2000)  & 1352  & 1495  & 1037  & 93123 (7884) \\
\hline
Total       &      & 0.000--0.168  & 227375 & 161424 & 8296 & 3538  & 2679 & 403314 (39567)
\enddata
\tablecomments{$^a$ Absolute magnitude limit for the sample. $^b$ Redshift limit for the sample. $^c$ The number of Q types in the sample. $^d$ The number of star-forming types in the sample. $^e$ The number of weak\,AGN types in the sample that have \oiii\ luminosity less than 10$^7$\,L$_{\sun}$. $^f$ The number of strong\,AGN type that have \oiii\ luminosity greater than 10$^7$\,L$_{\sun}$. $^g$ The number of QSO types in the sample. $^h$ The total number of galaxies in the sample. Numbers in parentheses are numbers of objects in each sample randomly selected for a calculation of a feature distance to the other type.
} 
\end{deluxetable*}

\section{Type assignment} \label{sec:assign}

\code{SpecObj.class} and \code{SpecObj.subclass}\footnote{\url{https://www.sdss3.org/dr8/spectro/catalogs.php}} are defined according to the following criteria:

\begin{enumerate}
    \item GALAXY: identified with a galaxy template; can have \code{SpecObj.subclass}: \begin{enumerate}
        \item STARFORMING: set based on whether the galaxy has detectable emission lines that are consistent with SF according to the criteria: 
        \begin{displaymath}
            \log_{10}(\oiii/\hb) < 0.7 - 1.2 (\log_{10}(\nii/\ha) + 0.4) 
        \end{displaymath}
        \item STARBURST: set if the galaxy is star-forming but has an equivalent width of
        H$\alpha$ greater than 50\,\AA
        \item AGN: set based on whether the galaxy has detectable emission lines that are consistent with being a Seyfert or LINER:
        \begin{displaymath}
            \log_{10}(\oiii/\hb) > 0.7 - 1.2 (\log_{10}(\nii/\ha) + 0.4) 
        \end{displaymath}
    \end{enumerate}
    \item QSO: identified with a QSO template.
\end{enumerate}

The \code{SpecObj.class} is assigned based on which type of redshifted PCA eigenspectra has a minimum $\chi^2$ value with respect to the observed specta.\footnote{\url{https://www.sdss3.org/dr8/algorithms/redshifts.php}}
If any galaxies or quasars have lines detected at the 10\,$\sigma$ level with $\sigma$\,$>$\,200\,km\,s$^{-1}$ at the 5\,$\sigma$ level, the indication ``BROADLINE'' is appended to their \code{SpecObj.subclass}. The range of trial redshifts used for the ``GALAXY'' class is $-$0.01 to 1.00, and 0.0333--7.00 is used for the ``QSO'' class.

\begin{figure*}
\newlength{\mylength}
\setlength{\mylength}{0.18\textwidth}
    % Row of Column Labels
    \gridline{
        \hspace{0.05\textwidth} % Minimal left margin
        \parbox[c]{\mylength}{\centering \textbf{Quiescent}}
        \parbox[c]{\mylength}{\centering \textbf{SF}}
        \parbox[c]{\mylength}{\centering \textbf{Weak AGN}}
        \parbox[c]{\mylength}{\centering \textbf{Strong AGN}}
        \parbox[c]{\mylength}{\centering \textbf{QSO}}
    }

    \vspace{-0.1cm} % Space between column labels and images

    % Image Rows with Row Labels
    \gridline{
        \raisebox{5\baselineskip}{\parbox[c]{0.05\textwidth}{\centering \textbf{180}}} % Centered row label with precise raise
        \fig{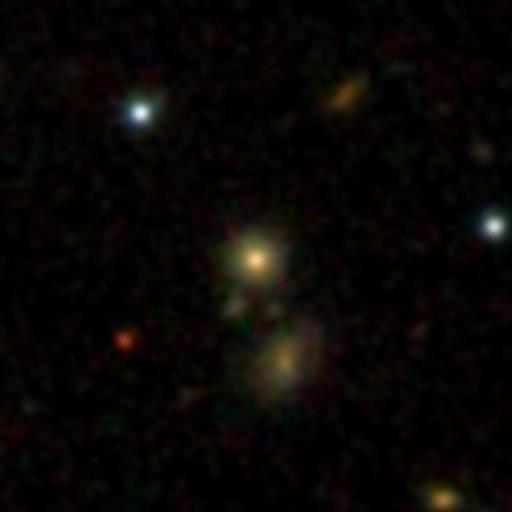}{\mylength}{}
        \fig{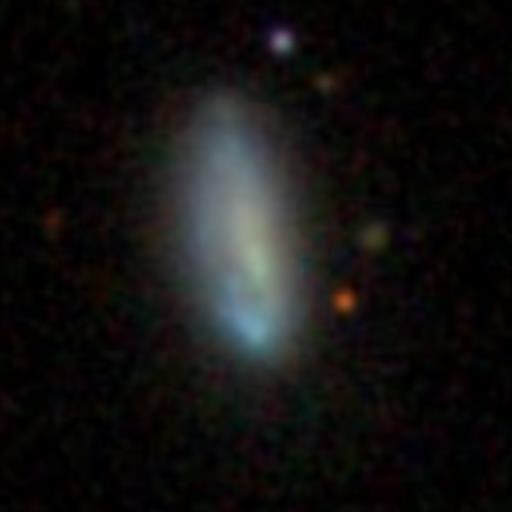}{\mylength}{}
        \fig{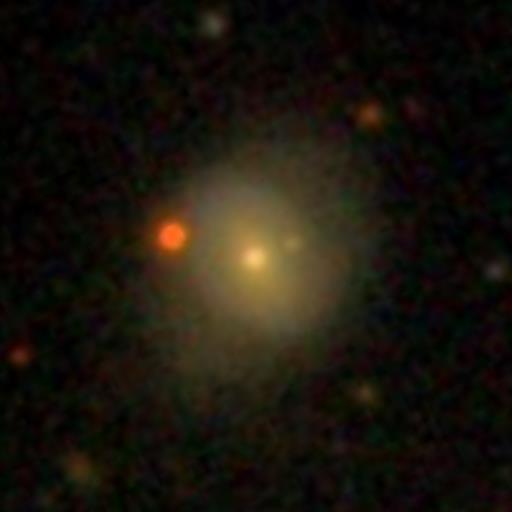}{\mylength}{}
        \fig{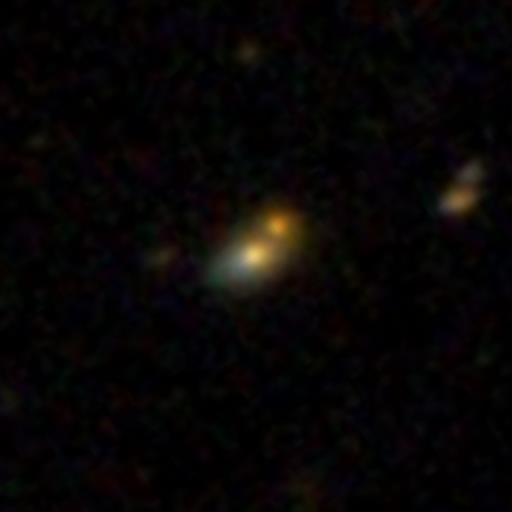}{\mylength}{}
        \fig{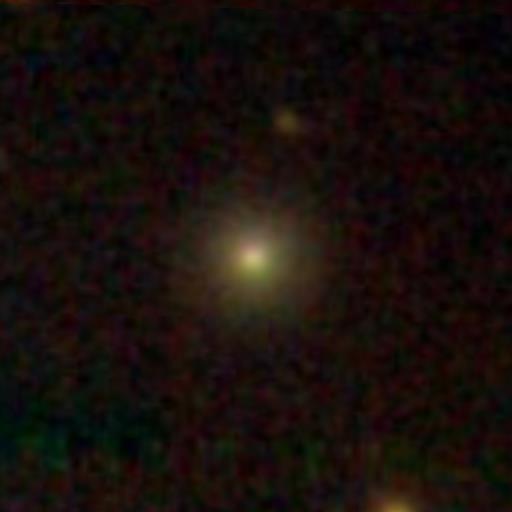}{\mylength}{}
    }
    \vspace{-1cm} % Reduce space between rows
    \gridline{
        \raisebox{5\baselineskip}{\parbox[c]{0.05\textwidth}{\centering \textbf{185}}}
        \fig{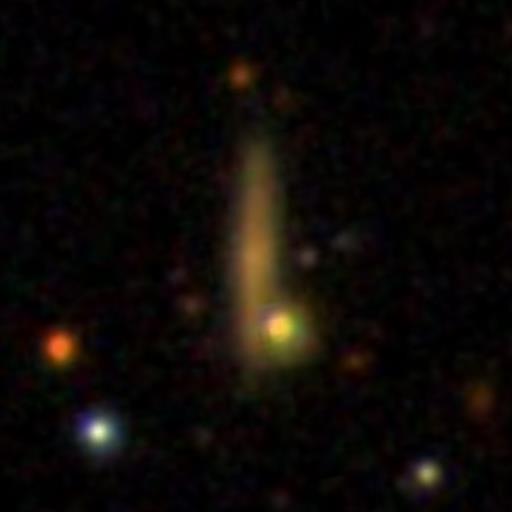}{\mylength}{}
        \fig{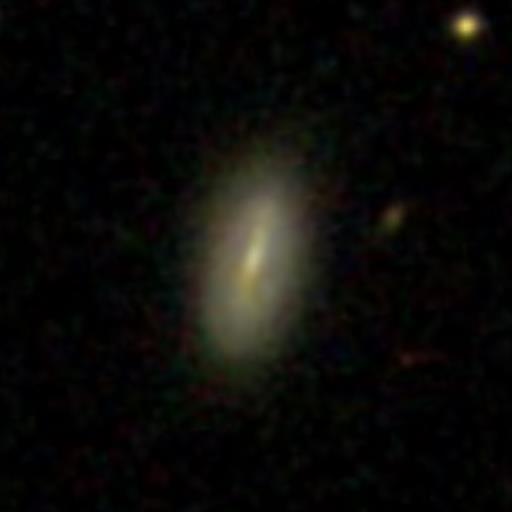}{\mylength}{}
        \fig{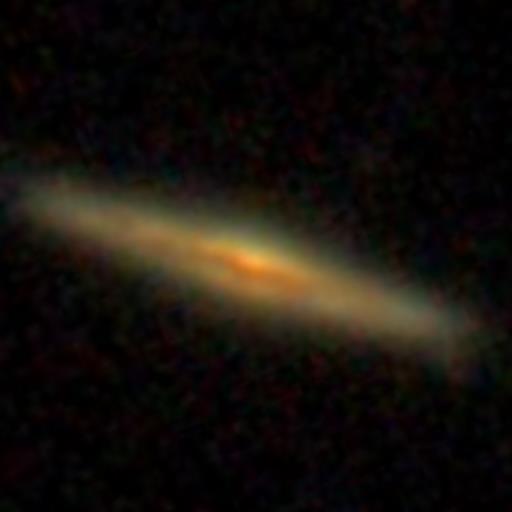}{\mylength}{}
        \fig{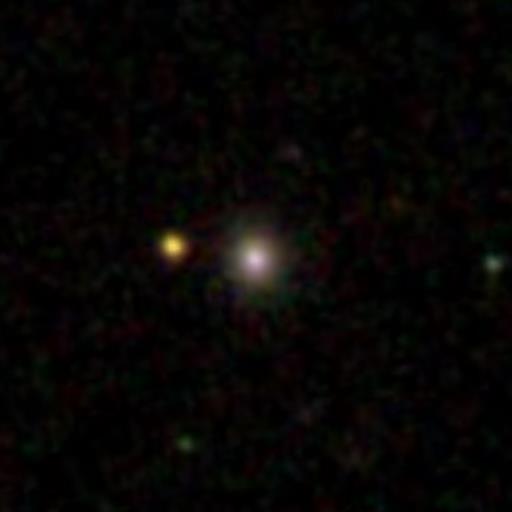}{\mylength}{}
        \fig{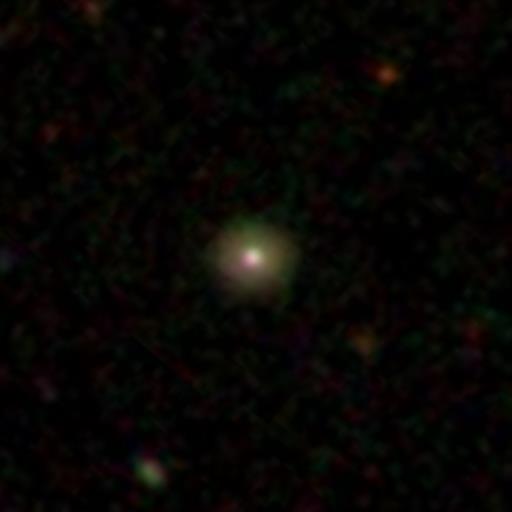}{\mylength}{}
    }
    \vspace{-1cm}
    \gridline{
        \raisebox{5\baselineskip}{\parbox[c]{0.05\textwidth}{\centering \textbf{190}}}
        \fig{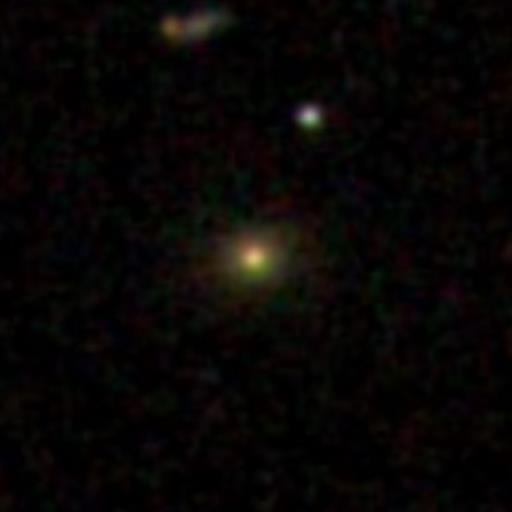}{\mylength}{}
        \fig{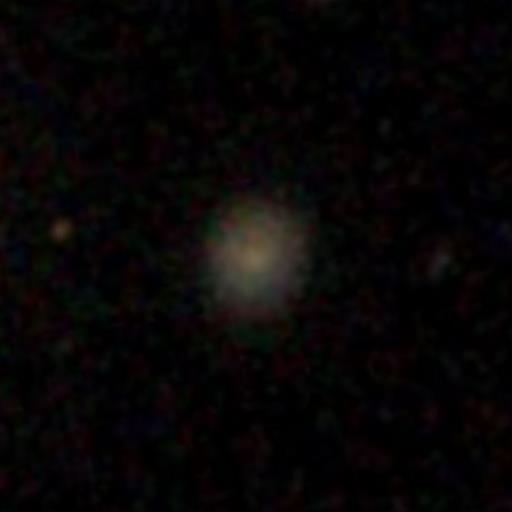}{\mylength}{}
        \fig{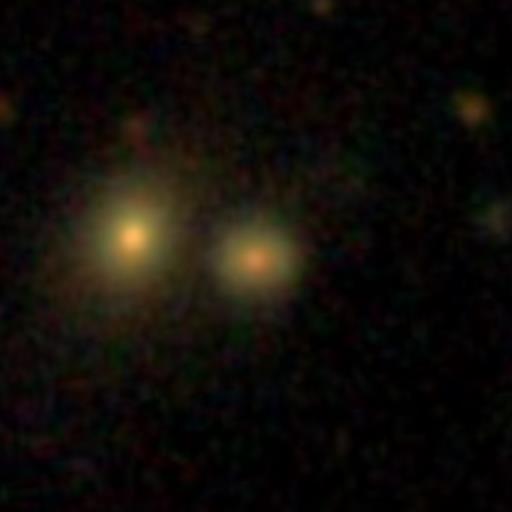}{\mylength}{}
        \fig{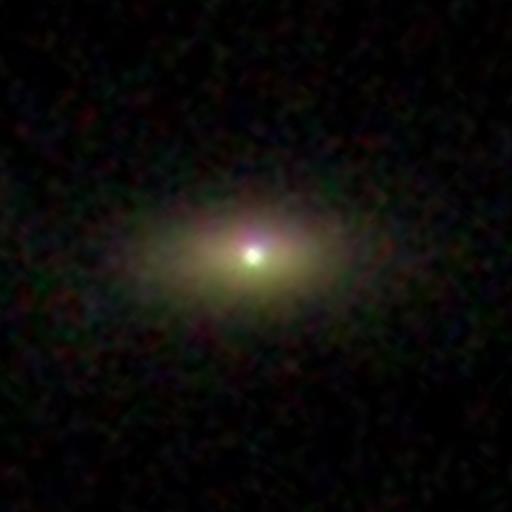}{\mylength}{}
        \fig{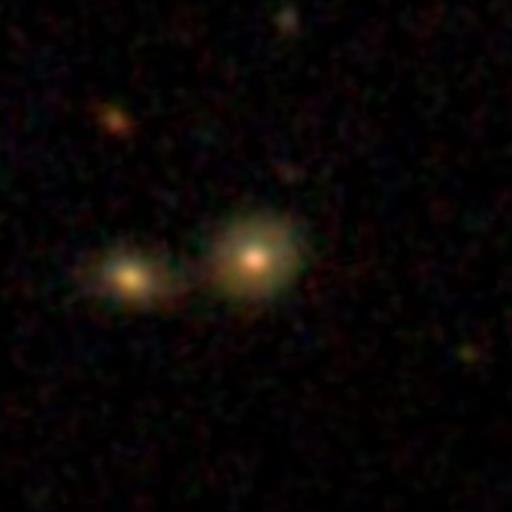}{\mylength}{}
    }
    \vspace{-1cm}
    \gridline{
        \raisebox{5\baselineskip}{\parbox[c]{0.05\textwidth}{\centering \textbf{195}}}
        \fig{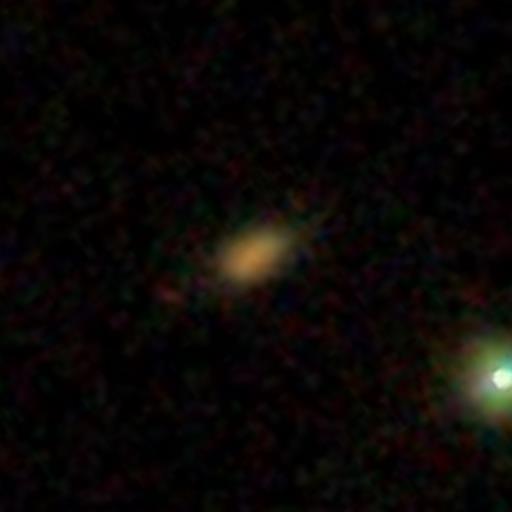}{\mylength}{}
        \fig{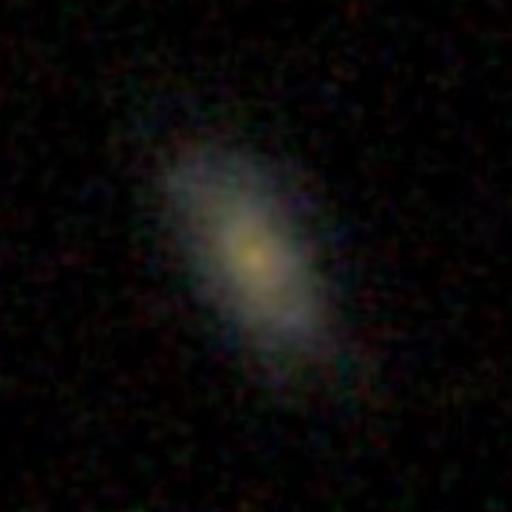}{\mylength}{}
        \fig{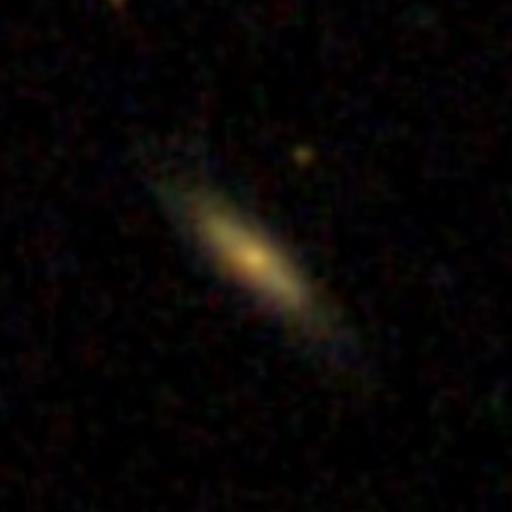}{\mylength}{}
        \fig{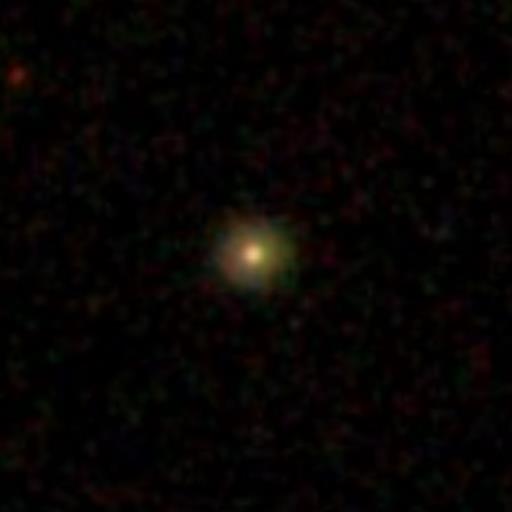}{\mylength}{}
        \fig{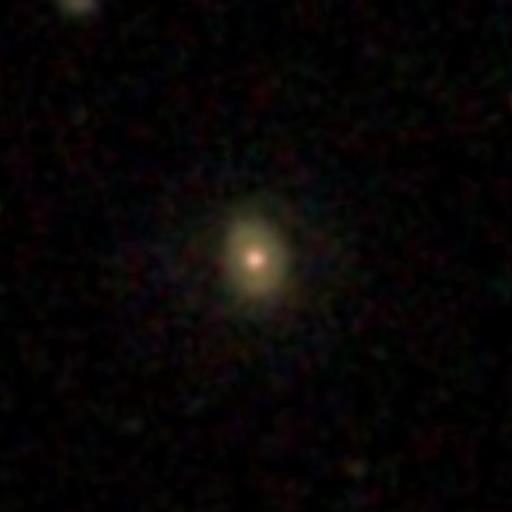}{\mylength}{}
    }
    \vspace{-1cm}
    \gridline{
        \raisebox{5\baselineskip}{\parbox[c]{0.05\textwidth}{\centering \textbf{200}}}
        \fig{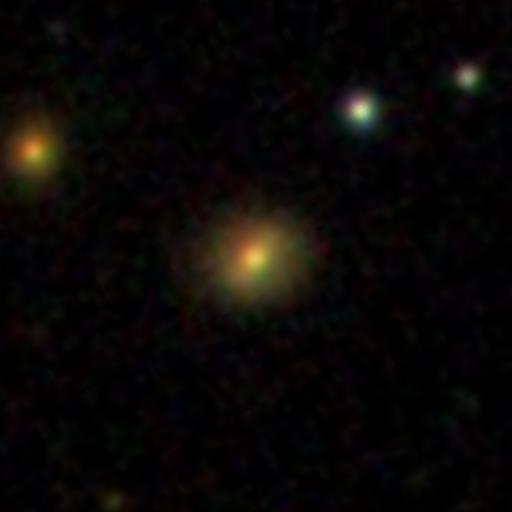}{\mylength}{}
        \fig{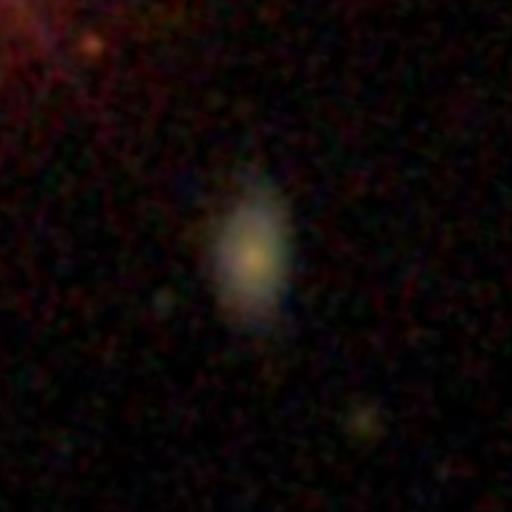}{\mylength}{}
        \fig{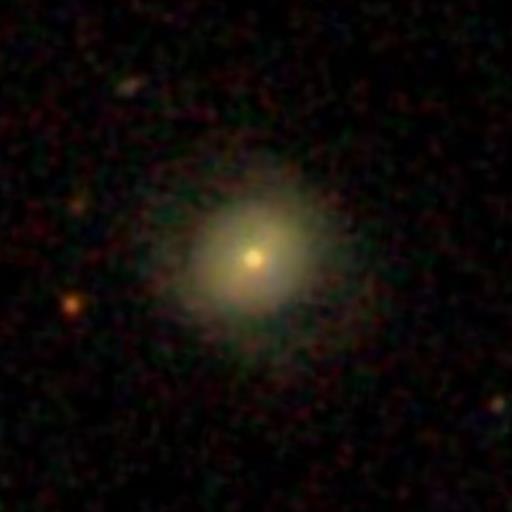}{\mylength}{}
        \fig{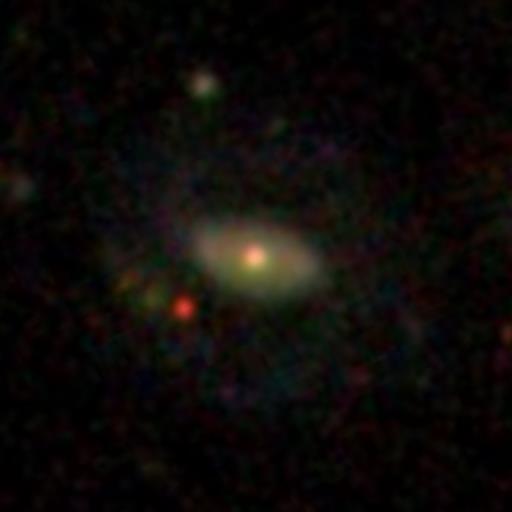}{\mylength}{}
        \fig{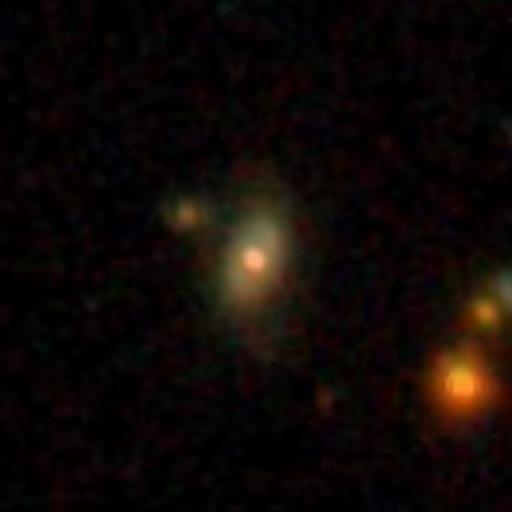}{\mylength}{}
    }
    \vspace{-1cm}
    \gridline{
        \raisebox{5\baselineskip}{\parbox[c]{0.05\textwidth}{\centering \textbf{205}}}
        \fig{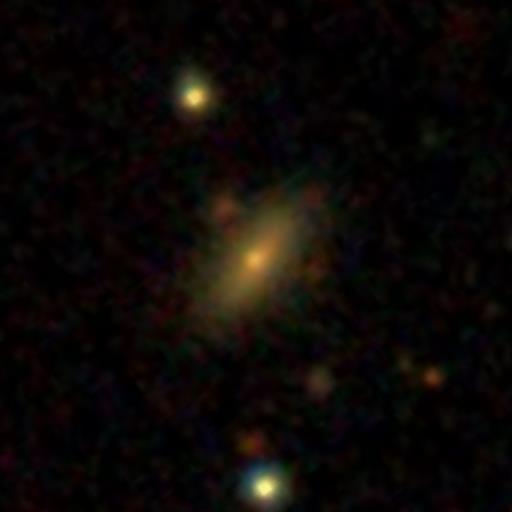}{\mylength}{}
        \fig{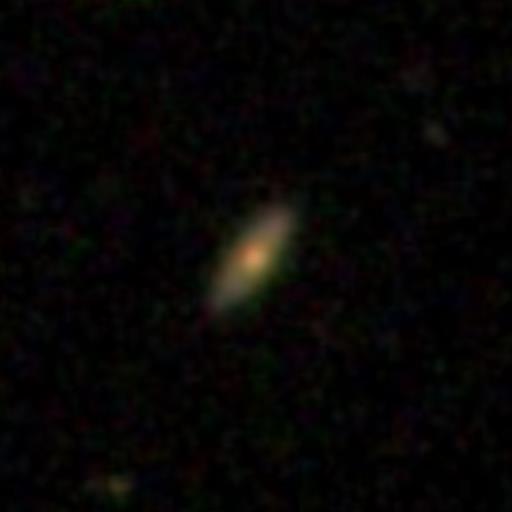}{\mylength}{}
        \fig{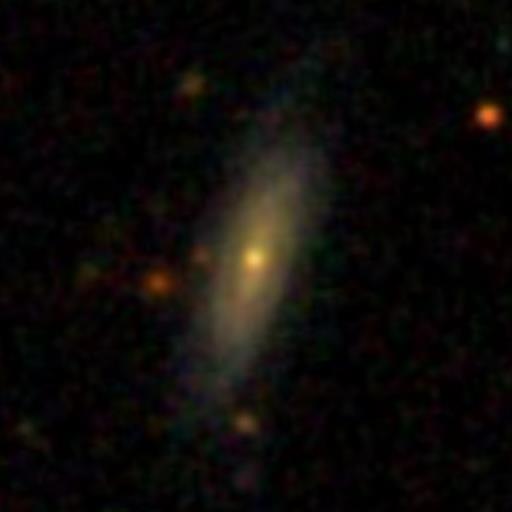}{\mylength}{}
        \fig{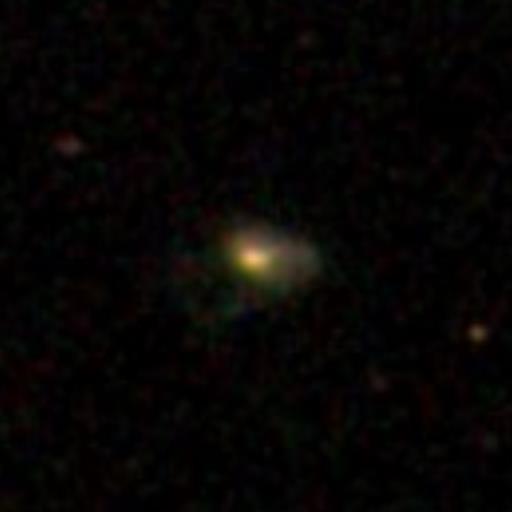}{\mylength}{}
        \fig{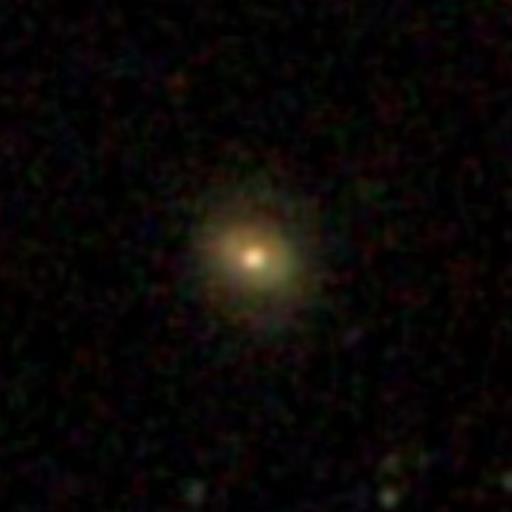}{\mylength}{}
    }
    \vspace{-1cm}
    \gridline{
        \raisebox{5\baselineskip}{\parbox[c]{0.05\textwidth}{\centering \textbf{210}}}
        \fig{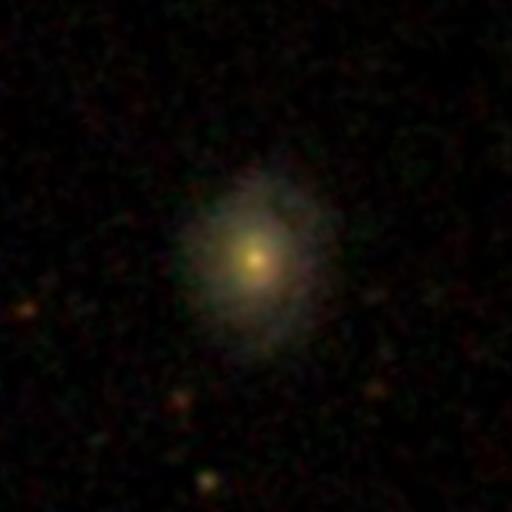}{\mylength}{}
        \fig{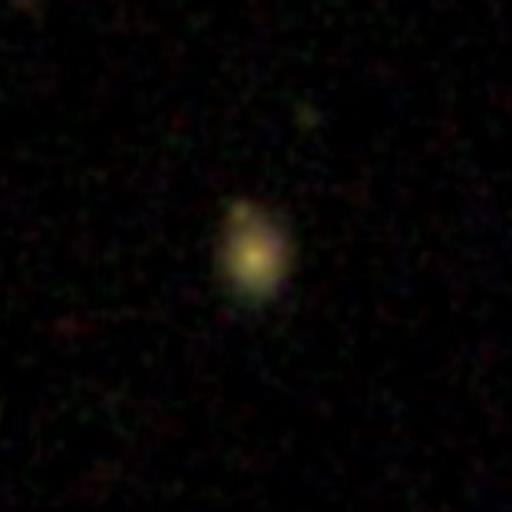}{\mylength}{}
        \fig{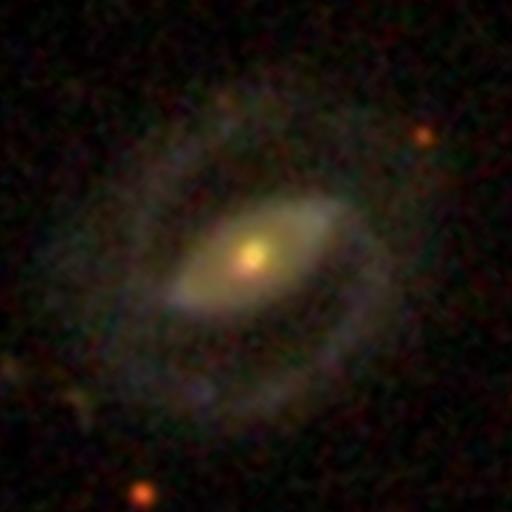}{\mylength}{}
        \fig{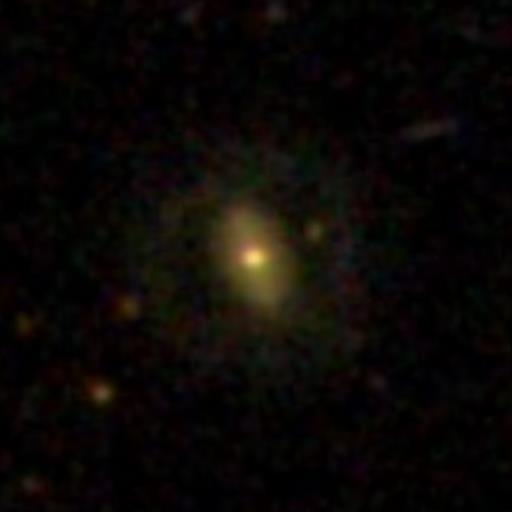}{\mylength}{}
        \fig{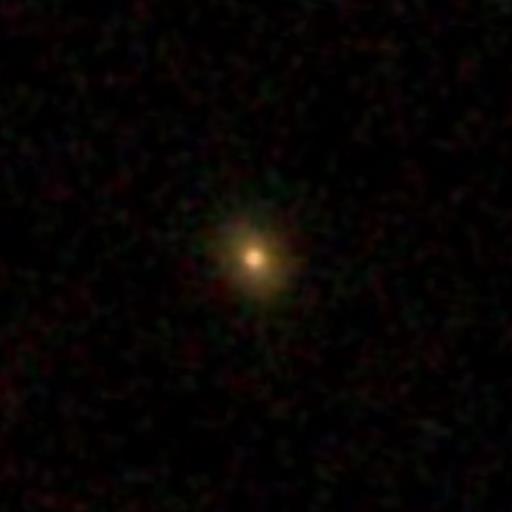}{\mylength}{}
    }
    \vspace{-0.5cm}
\caption{Image grid of various galaxy types across different classifications. Each column represents a distinct category: Q, star-forming (SF), weak\,AGN, strong\,AGN, and quasar (QSO). Each row represents different magnitude intervals labeled from 180 to 210 corresponding to $-$18\,$>$\,$M_r$\,$>$\,$-$18.5 and $-$21\,$>$\,$M_r$, respectively.\label{fig:cutout}}
\end{figure*}

Based on \code{SpecObj.class} and \code{SpecObj.subclass}, we assign a type to each galaxy as follows:

\begin{enumerate}
    \item Q: a sample with ``GALAXY'' \code{class} without any \code{subclass};
    \item SF: a sample with ``GALAXY'' \code{class} and any of ``STARFORMING,'' ``BROADLINE,'' ``STARBURST,'' ``STARBURST BROADLINE,'' ``STARFORMING BROADLINE'' \code{subclasses};
    \item Weak AGN: a sample with ``GALAXY'' \code{class} and ``AGN'' or ``AGN BROADLINE'' \code{subclasses} with $L_{\oiii} < 10^7 L_{\sun}$;
    \item Strong AGN: a sample with ``GALAXY'' \code{class} and ``AGN'' and ``AGN BROADLINE'' \code{subclasses} with $L_{\oiii} > 10^7 L_{\sun}$, and
    \item QSO: ``QSO'' \code{class},
\end{enumerate}

\noindent where $L_{\oiii}$ is the value we calculated using \code{galSpecLine.oiii\_5007\_flux} and \code{SpecObj.z}. The dividing value of weak and strong\,AGN $L_{\oiii}$\,$=$\,$10^7 L_{\sun}$ is adopted from \citet{Kauffmann03}. Figure~\ref{fig:cutout} shows exemplary images of galaxies randomly selected from the types above in the seven magnitude bins shown in Table~\ref{tab:sample}. We downloaded cutouts from SDSS SkyServer\footnote{\url{http://skyserver.sdss.org/dr16/}} with a box size of 0\arcmin.85$\times$0\arcmin.85, corresponding to 512$\times$512 pixels and a 0\arcsec.1\,pixel$^{-1}$ scale. The images in the top row are Q, SF, weak\,AGN, strong\,AGN, and QSO types with $-$18\,$>$\,$M_r$\,$>$\,$-$18.5. Subsequent rows indicate the same types for galaxies in increasingly faint magnitude bins.

\begin{figure}
 \gridline{ 
            \fig{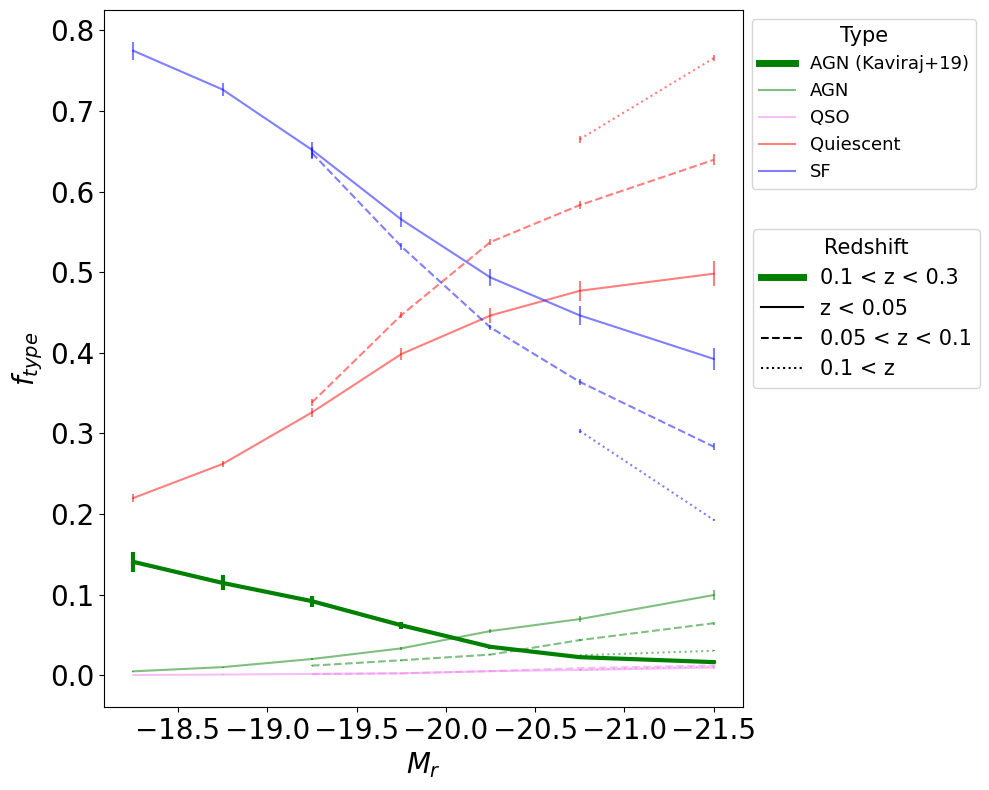}{0.5\textwidth}{}
          }
\caption{Fractions of AGN, QSO, Q, and SF types in a grid of $M_r$ and redshift bins. As galaxies get brighter, Q and AGN fractions increase, while the SF fraction decreases. As redshift increases, the fraction of Q increases, while AGN and SF fractions decrease. On the other hand, the AGN fraction from \citet{Kaviraj19} \emph{decreases} as galaxies get brighter. 
For calculating the effective AGN FB in Section~\ref{sec:dis}, we adopt the AGN fraction values from \citet{Kaviraj19} rather than \citetalias{Tempel14}, as the latter may be affected by flux and aperture limitations.
\label{fig:frac}}
\end{figure}

Figure~\ref{fig:frac} shows the fractional contribution for each galaxy type in bins of magnitude across different redshift ranges. The crossing fractions of Q versus SF can be understood as the color-magnitude relation \citep{Tully82}. 
In addition to our AGN-type fractions from \citetalias{Tempel14}, we present AGN fractions from \citet{Kaviraj19}, which show an opposite trend with magnitudes toward fainter magnitudes. The AGN types calculated from \citetalias{Tempel14} are classified based on the detectable emission lines \citep[such that $S/N$\,$>$\,3 in][]{Brinchmann04}, which are flux limited so the decreasing $f_{\mathrm{AGN}}$ in the low-luminosity end can be understood as a result of their low $S/N$. \citet{Kauffmann03} explain the increase $f_{\mathrm{AGN}}$ with the luminosities of galaxies as a result of the limitation of the 3\arcsec\ diameter fiber aperture, which collects photons more from the galactic central regions than from the outskirts.
In contrast, \citet{Kaviraj19} select AGN using an infrared color from the Wide-field Infrared Explorer \citep[WISE;][]{Wright10}, which exhibits higher $f_{\mathrm{AGN}}$ toward lower luminosity. We calculate $f_{\mathrm{AGN}}$ values from WISE photometry cross-matched with much-higher-angular resolution Hyper Suprime-Cam Subaru Stategic Program (HSC-SSP) data following \citet{Kaviraj19}, which is not limited by flux and aperture limitation, for the calculation of \emph{effective} AGN FB in Section~\ref{sec:dis}. As in \citet{Kaviraj19}, we adopt a search radius of 3\arcsec\ and require that only one HSC source is found within 6$^{\prime\prime}$ (approximately the size of the WISE point spread function). Some studies \citep[e.g.,][]{Lupi2020} have suggested that AGN fractions may be limited by contamination due to the poor resolution of ground-based infrared surveys. We test the impact of contamination by relaxing both the search radius and requirement of only a single HSC object within 6\arcsec. We find that either increasing the search radius or allowing additional nearby objects results in a modest decrease in $f_{\mathrm{AGN}}$ on average. Even under significantly relaxed requirements (search radius\,=\,6\arcsec, number of objects within 6\arcsec\,=\,5) we find that the same general trend remains, with $f_{\mathrm{AGN}}$ increasing toward fainter magnitudes, indicating that any contamination that may be present, even with our much more conservative requirements, does not result in spuriously high AGN fractions at the low-luminosity end.

In subsequent multidimensional analyses, we balance the numbers of galaxies of each type by randomly sampling from the Q and SF types to match the combined number of weak AGN, strong AGN, and QSOs in each sample (see the numbers in parentheses in Table~\ref{tab:sample}).

\section{Multidimensional analysis} \label{sec:clustering}

\subsection{\code{Orange}} \label{sec:orange}

We have used \code{Orange version 3.37 and 3.38}\footnote{\label{orange}\url{https://orangedatamining.com/}} for our multidimensional analyses. \code{Orange} is an open-source software for visual programming and interactive data visualization. \code{Orange} allows a user to upload catalog data and analyze them by applying commonly used machine learning algorithms simply by placing widgets on the canvas and connecting them.

\subsection{Feature parameter set} \label{sec:param}

We select the parameters used in our multidimensional analysis following \citet[][hereafter \citetalias{Banerji10}]{Banerji10}, who trained an artificial neural network on various sets of photometric inputs to determine the optimal combination for morphological classification into early types, spirals, and point sources/artifacts. \citetalias{Banerji10} found that a combination of profile fitting and adaptive weighted fitting parameters resulted in the best accuracy of more than 90\% compared with classifications from Galaxy Zoo. Similarly, we select additional parameters motivated by \citet{Gauci10}, who applied decision-tree learning algorithms and fuzzy inference systems to the photometric parameters of the SDSS DR7 catalog to distinguish between spiral galaxies, elliptical galaxies, or star/unknown galactic objects. 

We adopt the following set of parameters from the two studies, also listed in Table~\ref{tab:feat}:
\begin{enumerate}
    \item \emph{deVAB}: axial ratio of de Vaucouleurs profile fit model,
    \item \emph{expVAB}: axial ratio of exponential profile fit model,
    \item \emph{f(deV\_L)}: fractional likelihood of de Vaucouleurs fit,
    \item \emph{f(exp\_L)}: fractional likelihood of exponential disk fit,
    \item \emph{mRrCc}: the sum of second moments (extent),
    \item \emph{aE}: the sum of ellipticity components, and
    \item \emph{mCr4}: fourth-order moment (kurtosis).
\end{enumerate}
De Vaucouleurs and exponential parameters are derived from fits to to the two-dimensional profiles of each object in each band from the SDSS catalog.\footnote{\url{https://www.sdss3.org/dr8/algorithms/magnitudes.php\#mag_model}} The axial ratios and likelihoods for the two models are obtained in the fitting process. The likelihood function measures how well a statistical model explains the observed data. The fractional likelihoods are calculated as their fraction among the likelihoods for all models:\footnote{\url{https://www.sdss3.org/dr8/algorithms/classify.php\#photo_fits}} 
\begin{equation}
    f(deV\_L) = deV\_L / [deV\_L + exp\_L + star\_L].
\label{eq:frac}
\end{equation}
We refer to \citetalias{Banerji10} for the definitions of \emph{mRrCc}, \emph{aE}, and \emph{mCr4}, which are included in the second set of their input parameters. %\citetalias{Banerji10} resulted in the most accurate morphological classifications using a neural network when trained using both profile fitting and adaptive weighted scheme parameters.

We make use of the same set of parameters as used for classifying the morphology of galaxies using machine learning by \citetalias{Banerji10} to define distances in the morphological feature parameter space.

In order to construct the feature parameter set that represents exclusively galaxy properties, it is necessary to consider how the light from AGN affects each parameter. The distinctive colors of QSO are used for the detection mechanism \citep{Fan99,Kim15}, and as AGN become more dominant in bolometric luminosity, the spectral energy distribution tends to flatten, leading to increased uncertainties in photometric redshifts \citep{Chung14}. 
In addition, the presence of strong AGN emission lines in the SDSS filter bands, such as \hb, \oiii, and \ha, may influence their broadband colors. The central AGN point source can also increase the values of the concentration parameter \citep[see][]{Kauffmann03,Gabor09,Getachew22}. Therefore, we exclude the broadband colors and the concentration parameters from our main analysis and use them in a limited way for comparison purposes (see Figures~\ref{fig:dist_mag}--\ref{fig:dist_dens} and \ref{fig:dist_mag_separate}).  

\begin{figure}
 \gridline{ 
            \fig{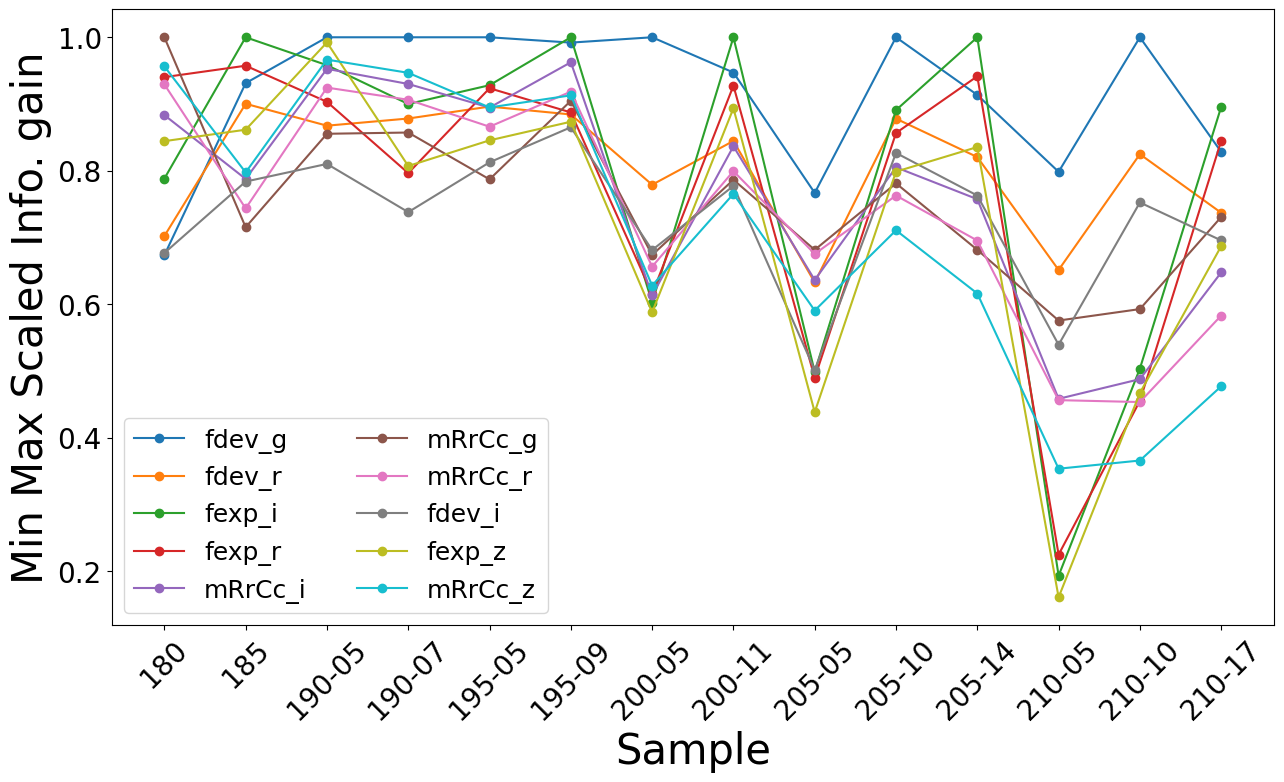}{0.5\textwidth}{}
          }
\caption{Min-max scaled Information Gain (IG) values of top 10 features across our samples. The legend shows the order of the average IG values across all samples (the left top feature has the largest IG values and right bottom feature has the least IG values). The fractional likelihood features had the most distinctive values in our spectral types.
\label{fig:info_gain}}
\end{figure}

We measure the significance of each feature using the \code{Rank} widget in \code{Orange}, which ranks the distinctiveness of each feature for the classification of the spectral type using the Information Gain (IG) scoring method. The IG scores indicate which morphological parameters are most significant for multidimensional analyses that follow. However, we do not attempt to classify the spectral types using this feature parameter set in this study. Higher IG scores are given to a feature that increases the opportunity to detect significant patterns and find informative splits in decision-tree algorithms. Figure~\ref{fig:info_gain} shows the IG values of the 10 features in Table~\ref{tab:feat} that have the highest average IG values. We normalized the scale of the IG values between 1 and 0 to eliminate the effect of the sample size. 
We find that the likelihoods for the galaxy light profiles are most distinctive among the spectral types, followed by the sum of second moments representing the extendedness.

\begin{deluxetable}{ll}
\tablecaption{Feature Parameter Set for \code{t-SNE} Analysis Adopted from \citetalias{Banerji10} and \citet{Gauci10} \label{tab:feat}}
\tablewidth{0pt}
\tablehead{
\colhead{Name} & \colhead{Description} 
}
\startdata
\emph{deVAB}                    & de Vaucouleurs fit axial ratio \\
\emph{expAB}                    & Exponential fit axial ratio \\
\emph{f(deV\_L)$^a$}            & de Vaucouleurs fit fractional likelihood \\
\emph{f(Exp\_L)$^a$}            & Exponential disk fit fractional likelihood \\
\emph{mRrCc}                    & Adaptive ($+$) shape measure \\
\emph{aE$^b$}                       & Adaptive ellipticity \\
\emph{mCr4}                     & Adaptive fourth moment \\
\enddata
\tablecomments{All features are for $u$, $g$, $r$, $i$, and $z$. $^a$ deV\_L / [deV\_L+exp\_L+star\_L], where deV\_L, exp\_L, and star\_L are log likelihoods for the de Vaucouleurs, exponential disk, and star fits (see Equation~\ref{eq:frac}). $^b$ $aE = 1 - \sqrt{\frac{1 - \sqrt{mE1^2 + mE2^2}}{1 + \sqrt{mE1^2 + mE2^2}}}$ which is from Equation~7 in \citetalias{Banerji10}, where $mE1$ and $mE2$ are adaptive E1 and E2 shape measurements (\url{https://skyserver.sdss.org/dr7/en/help/browser/browser.asp?n=PhotoObj\&t=U}).}
\end{deluxetable}

\subsection{\code{t-SNE}}

The \code{Orange} contains a \code{t-SNE} widget based on \citet{vanderMaaten08}. \code{t-SNE} reduces the number of dimensions of the data while preserving the relative distances between the data points. Unlike the linear-dimensionality reduction technique such as \code{PCA}, \code{t-SNE} captures nonlinear relationships in the data. \code{t-SNE} converts the high-dimensional Euclidean distances between data points into conditional probabilities that represent similarities. Then, it tries to optimize these probabilities in the lower-dimensional space. The algorithm minimizes the divergence between the probability distributions in the high- and low-dimensional spaces. The \code{t-SNE} can capture complex nonlinear relationships in the data, which makes it useful for visualizing clusters and patterns that are not apparent in the original high-dimensional space. 

We select `Normalize data' and `Apply PCA preprocessing' options for \code{Preprocessing} in the \code{t-SNE} widget. `Normalize data' normalizes each column by subtracting the column mean and dividing by the standard deviation before running PCA. `Apply PCA preprocessing' applies PCA preprocessing to speed up the algorithm and decorrelate the data. We leave `PCA Components' at its default value of 20. This is the number of principal components used when applying the PCA preprocessing. We also use \code{Parameters} set as default: (1) `PCA' is set for `Initialization' so that PCA positions the initial points along principal coordinate axes. (2) `Euclidean' is set for `Distance metric': the distance metric to be used when calculating distances between data points. (3) 30 is set for `Perplexity': the number of nearest neighbors over which distances will be preserved; using smaller values can reveal small, local clusters, while using large values tends to reveal the broader, global relationships between data points. (4) We leave blank the `Preserve global structure' check box. This option combines two different perplexity values (50 and 500) to try to preserve both the local and global structure. 5) The `Exaggeration' value is set as 1.00; the parameter increases the attractive forces between points and can directly be used to control the compactness of clusters. We chose the main value of each option as the default and changed it only if this resulted in the clustering on the t-SNE map being significantly improved. The \code{t-SNE} analysis is used for quantitative analysis in this study, while the quantitative analysis of \code{Distance} in Section~\ref{sec:dis} is independent of the  \code{t-SNE} analysis.

\begin{figure*}
\gridline{
        \hspace{0.1\textwidth} % Minimal left margin
        \parbox[c]{0.3\textwidth}{\centering \textbf{$z < 0.05$}}
        \parbox[c]{0.3\textwidth}{\centering \textbf{$0.05 < z < 0.1$}}
        \parbox[c]{0.3\textwidth}{\centering \textbf{$0.1 < z$}}
    }
 \gridline{ 
        \raisebox{5\baselineskip}{\parbox[c]{0.1\textwidth}{\centering \textbf{$-$21\,$>$\\$M_r$ }}} 
        \fig{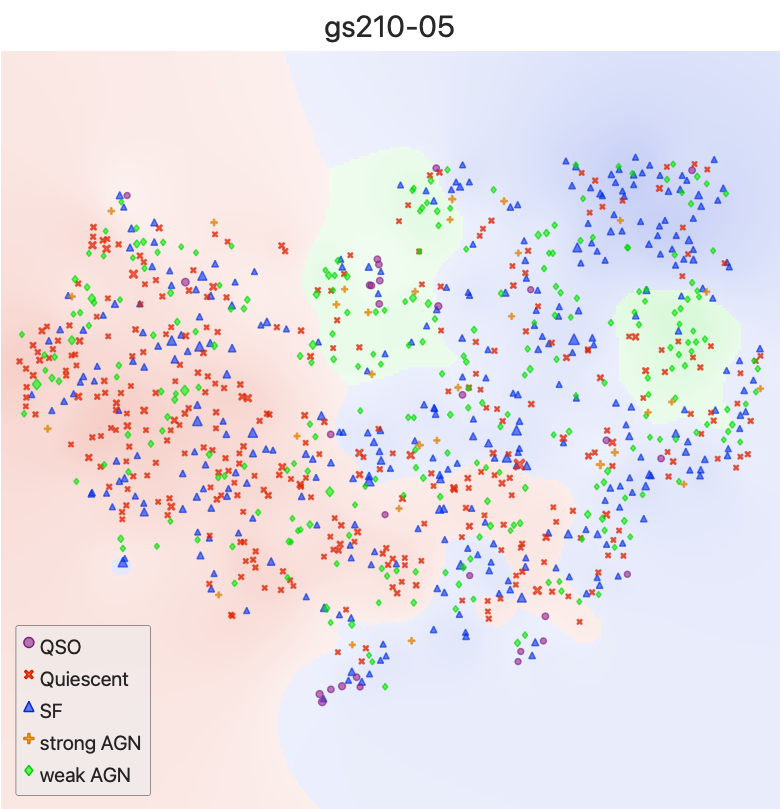}{0.3\textwidth}{}
          \fig{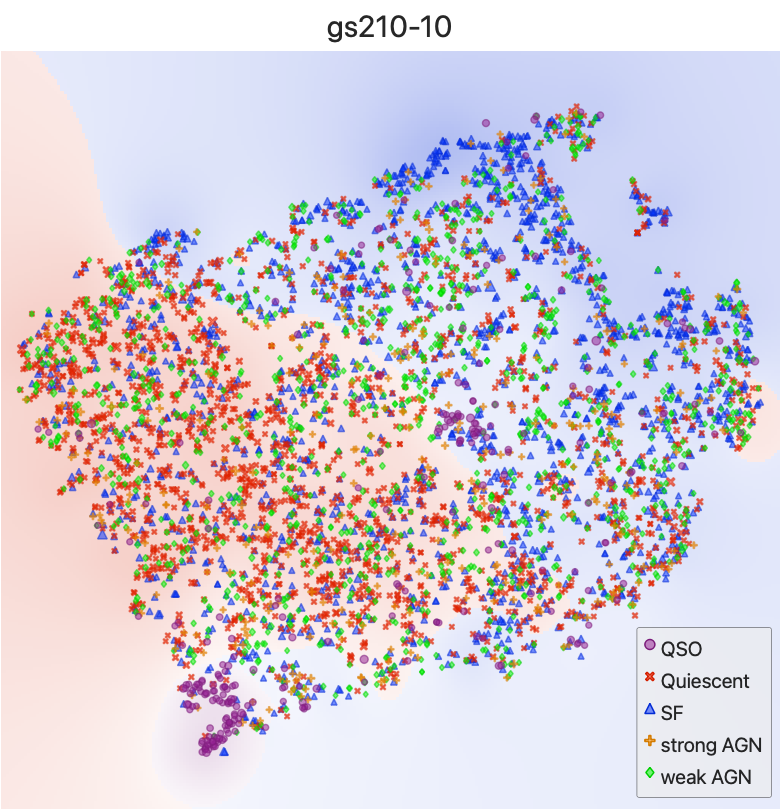}{0.3\textwidth}{}
          \fig{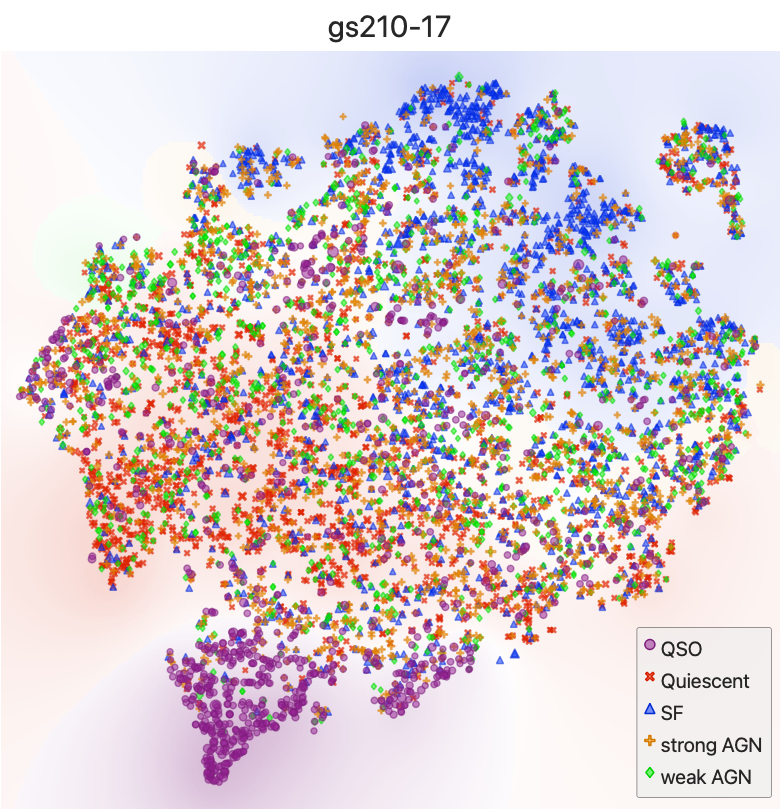}{0.3\textwidth}{}
          }
\gridline{
        \hspace{0.05\textwidth} % Minimal left margin
        \parbox[c]{\mylength}{\centering \textbf{$-$20.5\,$>$\,$M_r$\,$>$\,$-$21}}
    }
\vspace{-1cm}
\gridline{ 
        \raisebox{5\baselineskip}{\parbox[c]{0.1\textwidth}{\centering \textbf{$-$20.5\,$>$\\$M_r$\\$>$\,$-$21}}} 
        \fig{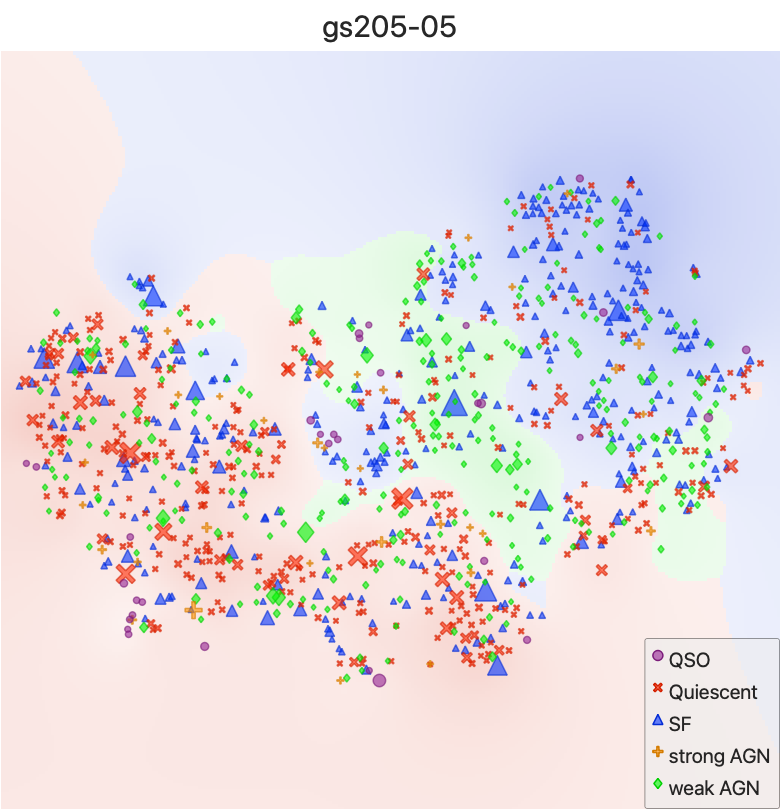}{0.3\textwidth}{}
          \fig{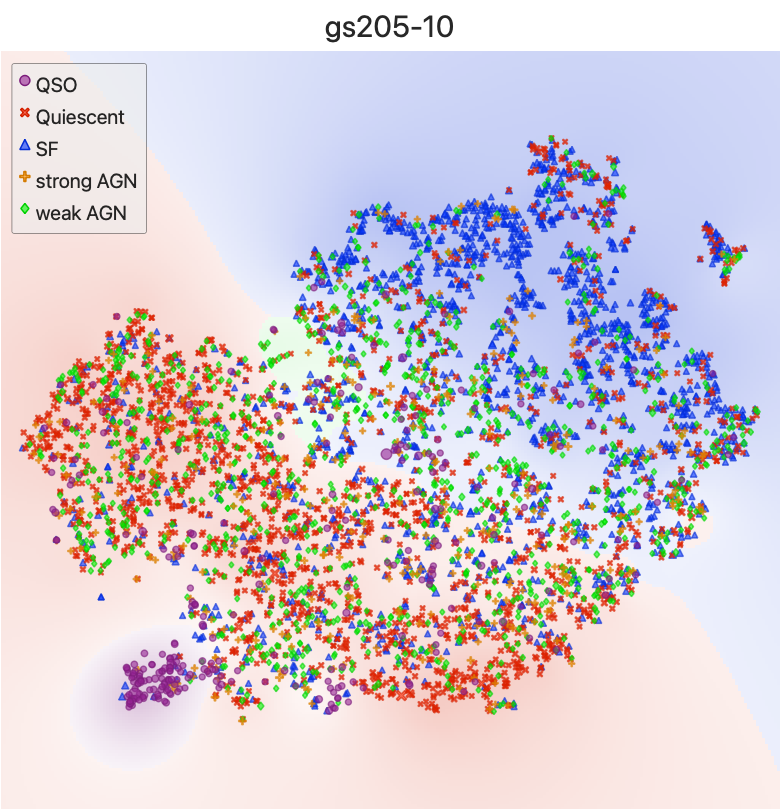}{0.3\textwidth}{}
          \fig{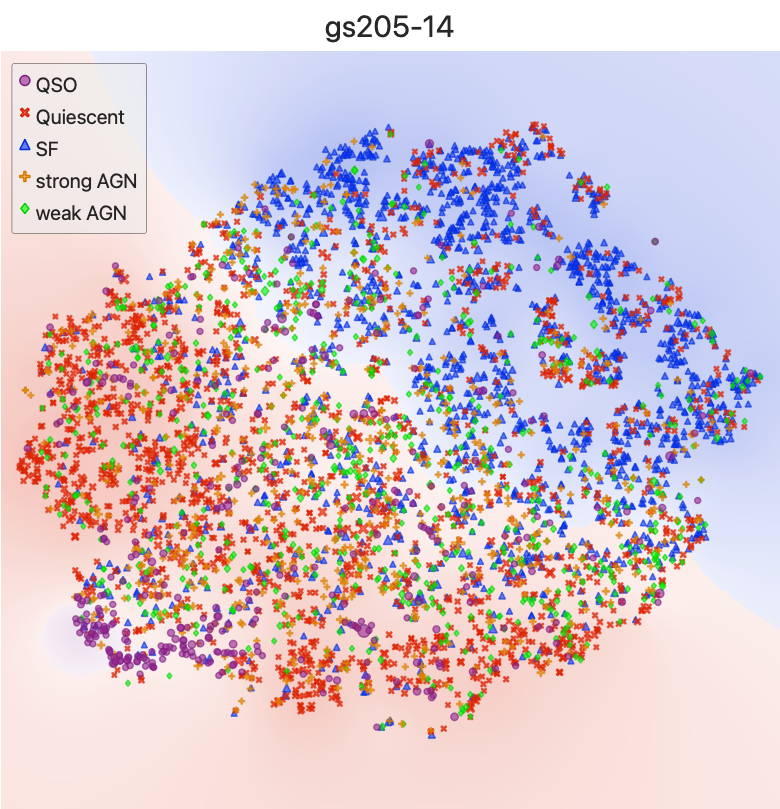}{0.3\textwidth}{}
          }
 \gridline{ \raisebox{5\baselineskip}{\parbox[c]{0.1\textwidth}{\centering \textbf{$-$20\,$>$\\$M_r$\\$>$\,$-$20.5}}} 
 \leftfig{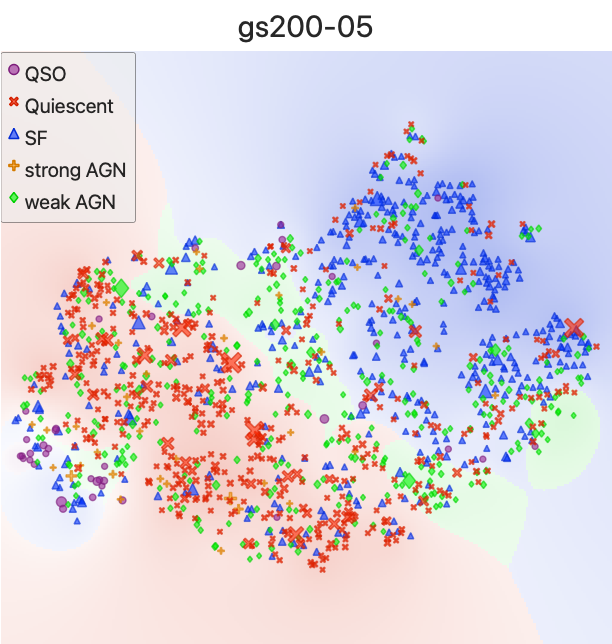}{0.3\textwidth}{}
          \leftfig{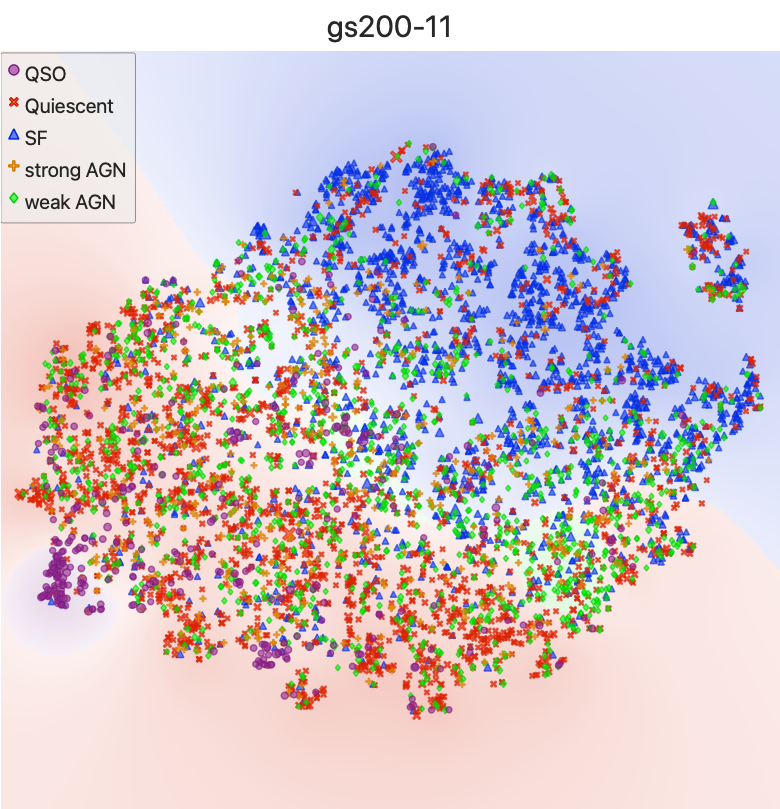}{0.3\textwidth}{}
          }
\caption{\code{t-SNE} result for samples brighter than $M_r$\,$=$\,$-21$ at the top, $-$21\,$>$\,$M_r$\,$>$\,$-$20.5 in the middle, and $-$20.5\,$>$\,$M_r$\,$>$\,$-$20 on the bottom. Samples with a redshift less than 0.05 are in the first column, 0.05\,$<$\,$z$\,$<$\,0.1 are in the second column, and those larger than 0.1 are in the third column. QSO, Q, SF, strong\,AGN, and weak\,AGN types are represented as a violet circle, red ``x'', blue triangle, orange cross, and green diamond, respectively.
\label{fig:tsne_low1}}
\end{figure*}

\begin{figure*}
\gridline{
        \hspace{0.1\textwidth} % Minimal left margin
        \parbox[c]{0.3\textwidth}{\centering \textbf{$z < 0.05$}}
        \parbox[c]{0.6\textwidth}{\centering \textbf{$0.05 < z < 0.1$}}
    }
 \gridline{ 
 \raisebox{5\baselineskip}{\parbox[c]{0.1\textwidth}{\centering \textbf{$-$19.5\,$>$\\$M_r$\\$>$\,$-$20 }}} 
            \leftfig{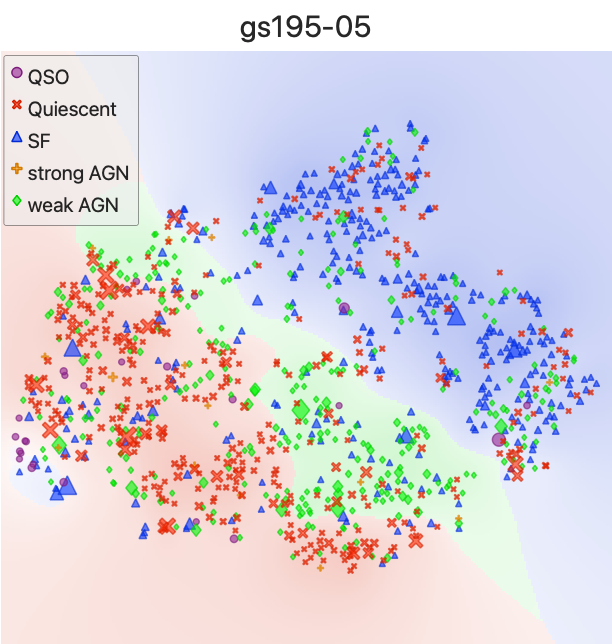}{0.3\textwidth}{}
            \leftfig{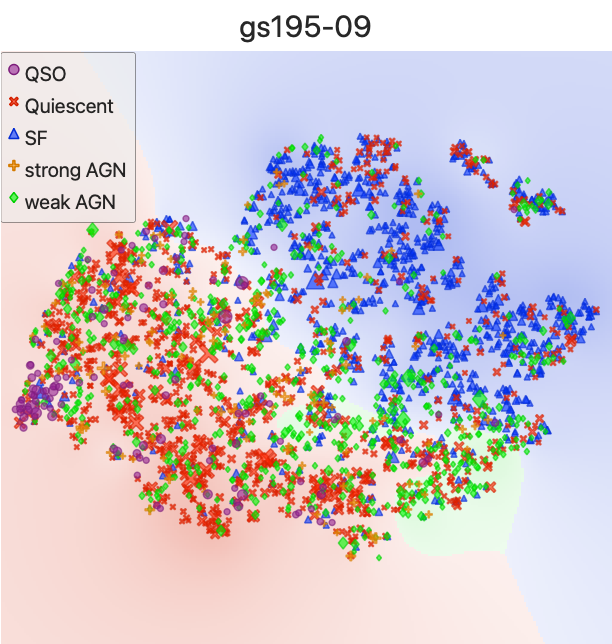}{0.3\textwidth}{}
          }
\vspace{-1cm}
 \gridline{ 
 \raisebox{5\baselineskip}{\parbox[c]{0.1\textwidth}{\centering \textbf{$-$19\,$>$\\$M_r$\\$>$\,$-$19.5 }}} 
            \leftfig{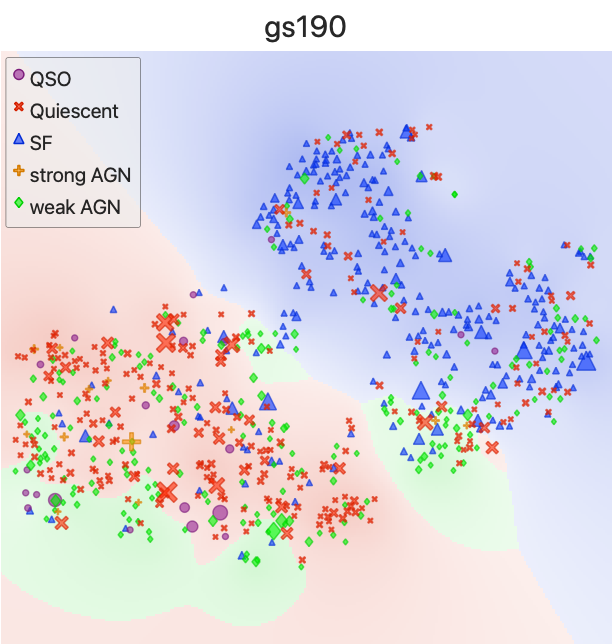}{0.3\textwidth}{}
            \leftfig{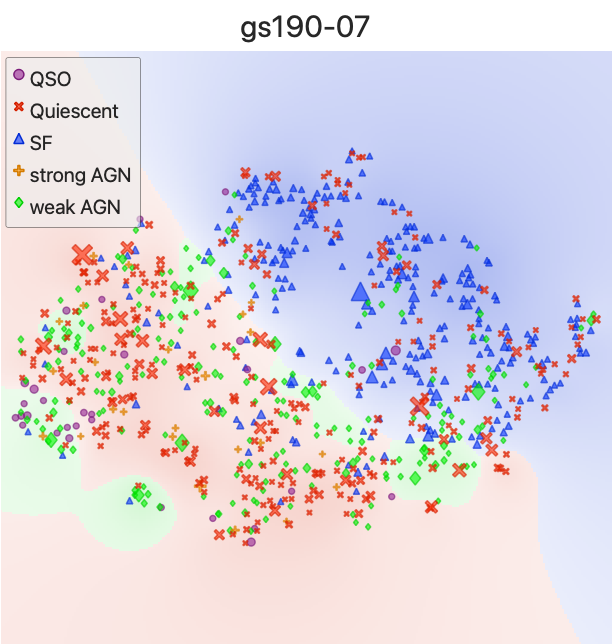}{0.3\textwidth}{}
          }
\vspace{-1cm}
 \gridline{ 
 \raisebox{5\baselineskip}{\parbox[c]{0.1\textwidth}{\centering \textbf{$-$18.5\,$>$\\$M_r$\\$>$\,$-$19 }}}
            \leftfig{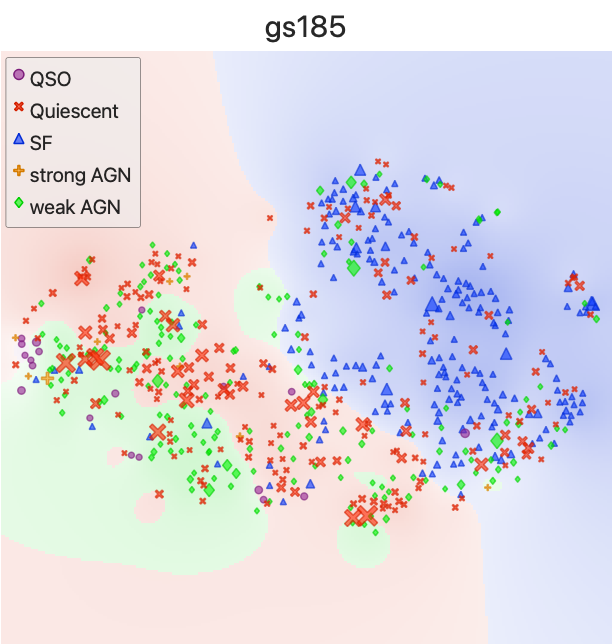}{0.3\textwidth}{}
          }
\vspace{-1cm}
 \gridline{ 
 \raisebox{5\baselineskip}{\parbox[c]{0.1\textwidth}{\centering \textbf{$-$18\,$>$\\$M_r$\\$>$\,$-$18.5}}}
            \leftfig{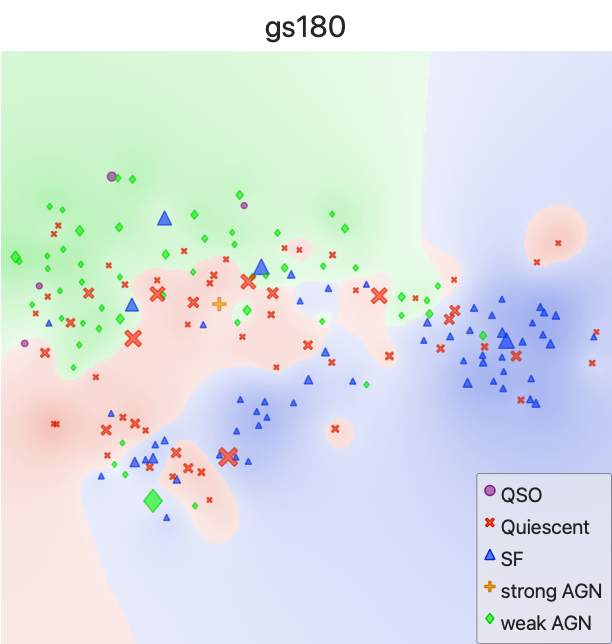}{0.3\textwidth}{}
          }
\caption{As in Figure~\ref{fig:tsne_low1}, but now for samples with $-$19.5\,$>$\,$M_r$\,$>$\,$-$20 in the first row, $-$19.5\,$>$\,$M_r$\,$>$\,$-$19 in the second row, and $-$19\,$>$\,$M_r$\,$>$\,$-$18.5 in the third row, and $-$18.5\,$>$\,$M_r$\,$>$\,$-$18 in the fourth row.
\label{fig:tsne_low2}}
\end{figure*}

Figures~\ref{fig:tsne_low1}--\ref{fig:tsne_low2} show the results of our \code{t-SNE} analysis for samples in Table~\ref{tab:sample} ordered by absolute magnitude $M_r$ from top (brightest) to bottom (faintest) and the redshift increasing from left to right. The points clustered together have similar values to the morphological characteristic parameters shown in Table~\ref{tab:feat}. Symbols and colors are assigned differently for each spectral type of galaxy. The size of the symbol represents the normalized environmental density of the galaxy for a smoothing scale a\,$=$\,1\,$h^{-1}$\,Mpc from \citetalias{Tempel14}. AGN types appear in the transition region between SF and Q types, while the QSO-type clusters at the corners of the Q type. Furthermore, the green region (weak\,AGN) appears to be associated with the red (Q) region in the dim sample (bottom of Figure~\ref{fig:tsne_low2}) but gradually moves toward the blue (SF) region as the sample becomes brighter.

\subsection{Distance}

The morphological similarity between galaxies can be measured quantitatively with distances in the multiparameter space. We use the \code{Orange} widgets \code{Distance} and \code{Distance Transformation} to calculate the distances between types in each sample in the normalized multiparameter space. We use the following three distance metrics:
\begin{enumerate}
    \item Euclidean: the straight-line distances between two points in a space, calculated using the Pythagorean formula, the square root of the sum of squared differences between corresponding coordinates;
    \item Manhattan: (also called taxicab or city block distance) measures the sum of absolute differences between coordinates;
    \item Cosine: measures the angular difference between two vectors, regardless of their magnitudes.
\end{enumerate}

\begin{figure*}
 \gridline{ 
            \fig{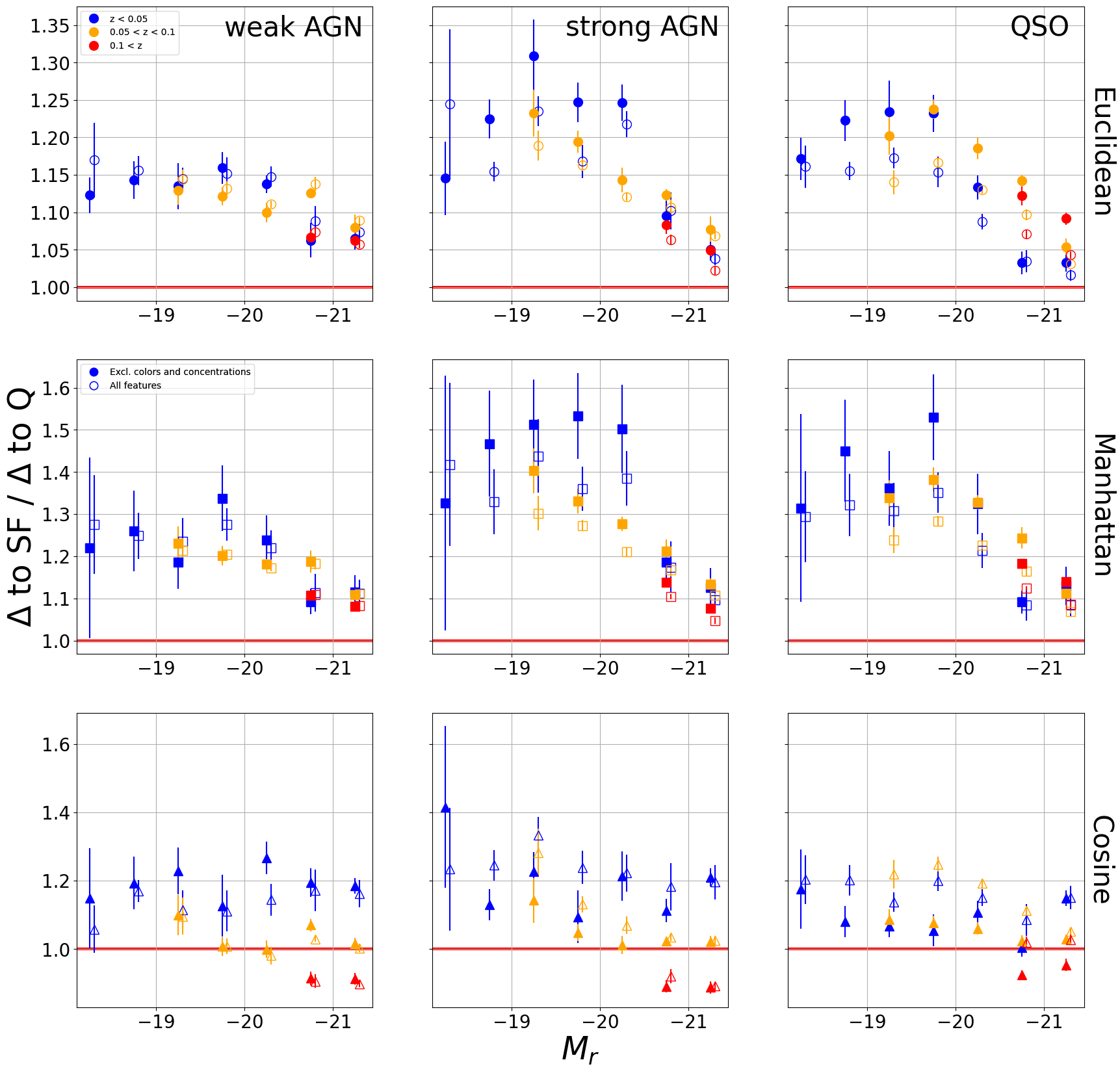}{\textwidth}{}
          }
\caption{Distances to SF ($\Delta$\,to\,SF) normalized by distances to Q ($\Delta$\,to\,Q) types from weak\,AGN (left), strong\,AGN (center), and QSO (right) types, in Euclidean (top), Manhattan (middle), and Cosine (bottom) distance metrics, as a function of absolute magnitude in $r$ band $M_r$. The color of a symbol represents the redshift range of a sample. Solid circles are distances on the morphological feature space in Table~\ref{tab:feat}. The open circles are distances measured adding colors and concentrations. AGN and QSO types have closer distances to Q than SF type, especially for magnitudes $-19 > M_r > -20$.
\label{fig:dist_mag}}
\end{figure*}

\begin{figure*}
 \gridline{ 
            \fig{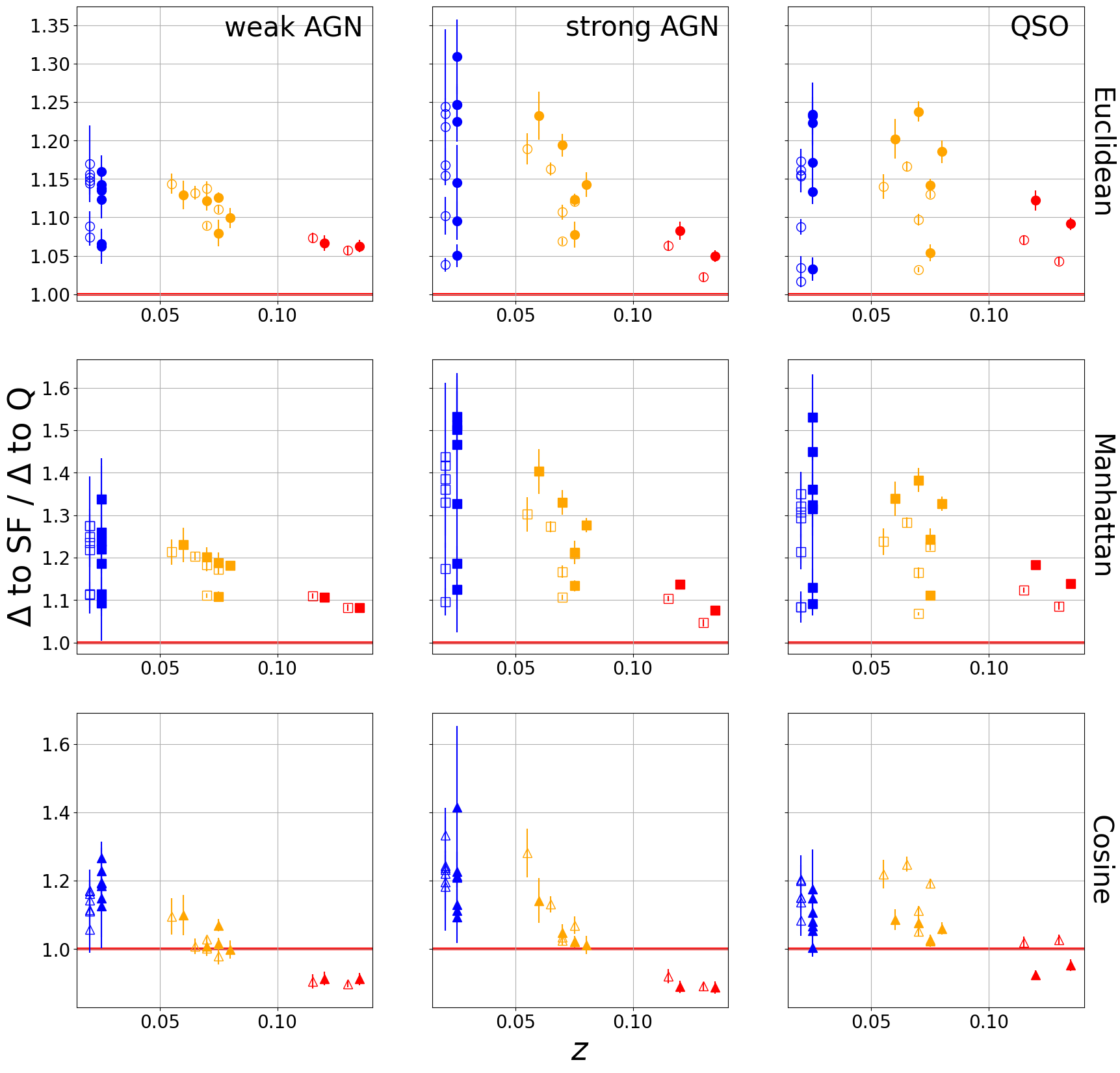}{\textwidth}{}
          }
\caption{As in Figure~\ref{fig:dist_mag} but now as a function of the redshift.
\label{fig:dist_z}}
\end{figure*}

\begin{figure*}
 \gridline{ 
            \fig{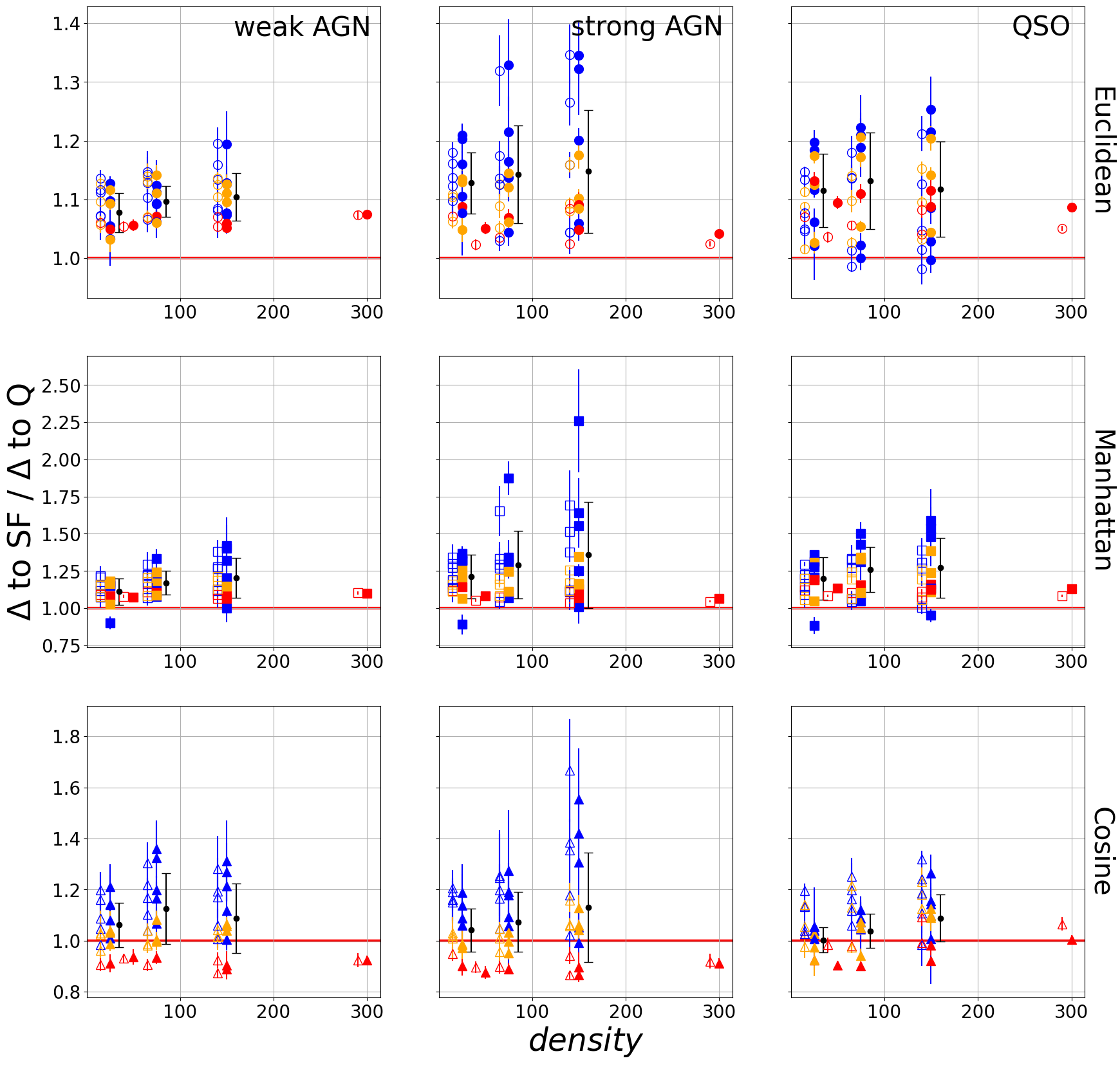}{\textwidth}{}
          }
\caption{As in Figure~\ref{fig:dist_mag} but now as a function of the normalized environmental density of a galaxy with a smoothing scale (a\,$=$\,1\,$h^{-1}$\,Mpc) from \citetalias{Tempel14}. 
\label{fig:dist_dens}}
\end{figure*}

Figures~\ref{fig:dist_mag}--\ref{fig:dist_dens} show the distance ratios between the SF and Q types from the weak\,AGN (left), strong\,AGN (center) and QSO (right) types in Euclidean (top), Manhattan (middle), and Cosine (bottom) distance metrics as functions of magnitude, redshift, and density, respectively.  In order to avoid biases as a result of different sample sizes, we selected 10 random samples with the same number of Q and SF types and measured the distances to AGN and QSO types (see the numbers in parentheses in the fourth and fifth columns of Table~\ref{tab:sample}), taking the average and standard deviation.

We obtain qualitative agreement across distance metrics, with all AGN and QSO types having distances closer to the Q than the SF type ($\Delta$\,to\,SF\,/\,$\Delta$\,to\,Q\,$>$\,1), except for the highest redshift samples in Cosine metric. In particular, the strong\,AGN resembles Q, larger $\Delta$\,to\,SF\,/\,$\Delta$\,to\,Q, than the weak\,AGN \citep[see][]{Kauffmann03}, and the ratio appears to be greatest at around $-19$\,$>$\,M$_{r}$\,$>$\,$-20$ and decreases as the magnitude decreases.
In Figure~\ref{fig:dist_z}, the scatter and the value of $\Delta$\,to\,SF\,/\,$\Delta$\,to\,Q itself tend to decrease as the redshift increases. This can be viewed as a marginal distribution of Figure~\ref{fig:dist_mag} as a function of the redshift (different colors) where the lower-redshift sample (blue) includes wide ranges of magnitudes, while the high-redshift sample (red) covers only bright galaxies with low-$\Delta$-to-SF/$\Delta$-to-Q values.
The weak relationship between the $\Delta$-to-SF/$\Delta$-to-Q values of AGN and the environmental density is displayed in Figure~\ref{fig:dist_dens} having greater $\Delta$-to-SF/$\Delta$-to-Q values in the high-density region except for the highest bin with a value of around 300.

For each spectral type, we analyzed the probability distribution of morphological classifications spanning early-type galaxies (E, S0) to late-type spirals (Sab, Scd). The \citetalias{Tempel14} catalog contains the probabilities derived by \citet{Huertas-Company11}. \citet{Huertas-Company11} trained support vector machines using three types of parameters: (1) $k$-corrected $g-r$ and $r-i$ colors; (2) axis ratios (ISO$_i$, deVAB$_i$); and (3) light concentration ($R_{90}$/$R_{50}$ in the $i$ band) to associate a probability value to each galaxy instead of a binary classification. 

\begin{figure}
 \gridline{ 
            \fig{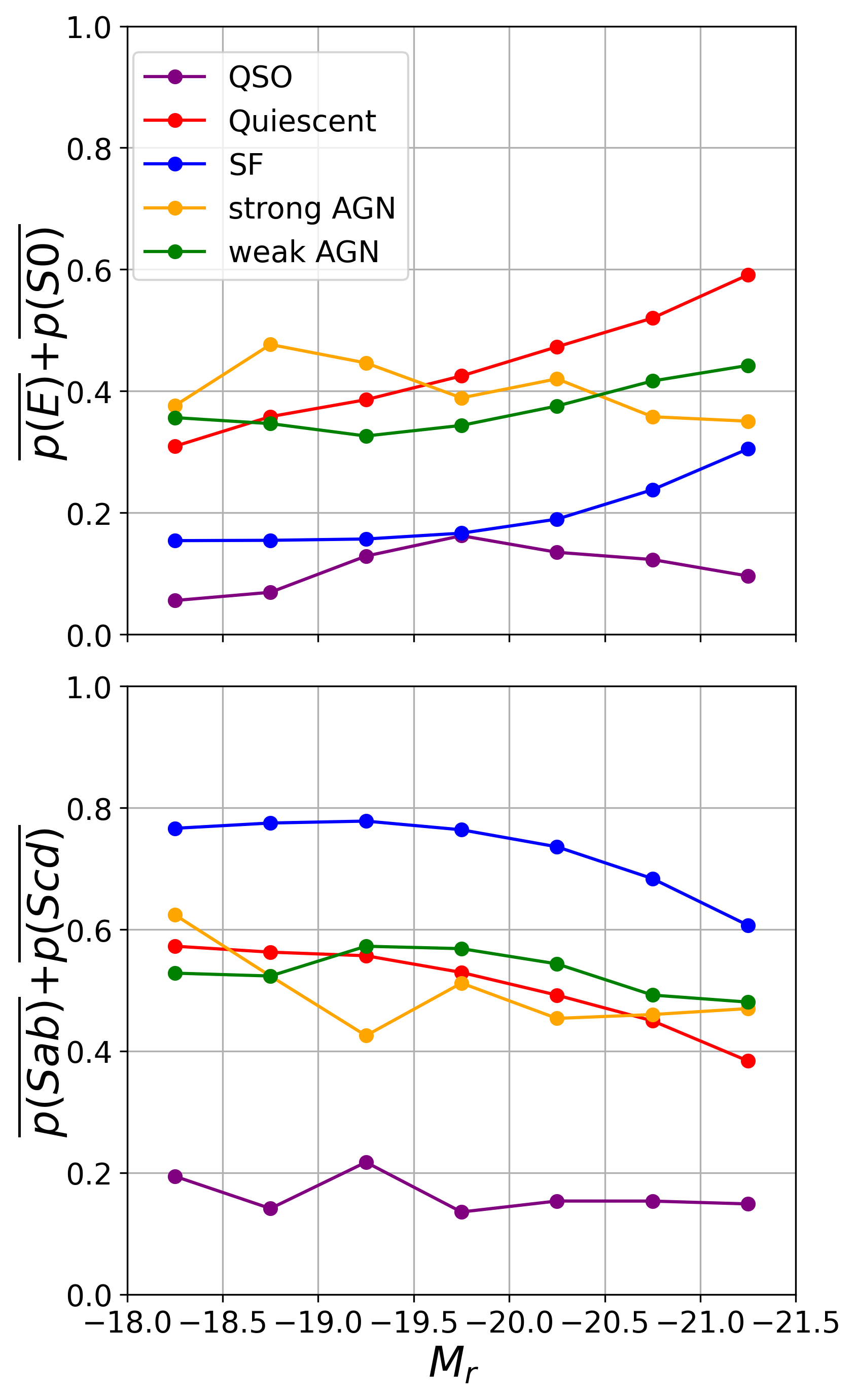}{0.5\textwidth}{}
          }
\caption{The sums of average probability that galaxies in our spectral type are E or S0 (upper) or Sab or Scd (lower) Hubble type. The probabilities are from the catalog of \citetalias{Tempel14}, originally derived by \citet{Huertas-Company11}. On average, weak and strong\,AGN types have similar probabilities to the Q type than the SF type, which agrees with our distance analysis result. In contrast, the QSO type has the lowest probability values of being any of the Hubble types. 
\label{fig:prob}}
\end{figure}

Figure~\ref{fig:prob} shows the summed mean probability of galaxies in each type being early type or S0 (upper) and Sab or Scd (lower). Both Q and SF types have an increasing probability of being early type or S0 as their luminosities increase, whereas the probabilities of being Sab or Scd decrease as their luminosities increase. On the other hand, weak and strong AGN types do not show any distinctive trend, probably because of the smaller number of galaxies in the samples. Nevertheless, AGN types have similar probability ranges as the Q type. SF types have a lower probability of being early type or S0 and a larger probability of being an Sab or Scd type when compared with Q and both types of AGN. On the bright end, though, $\overline{p(E)}$+$\overline{p(S0)}$ of the SF type is getting closer to those of AGN types, while the Q type is diverging. This trend is similar to our distance result.

\begin{figure}
 \gridline{ 
            \fig{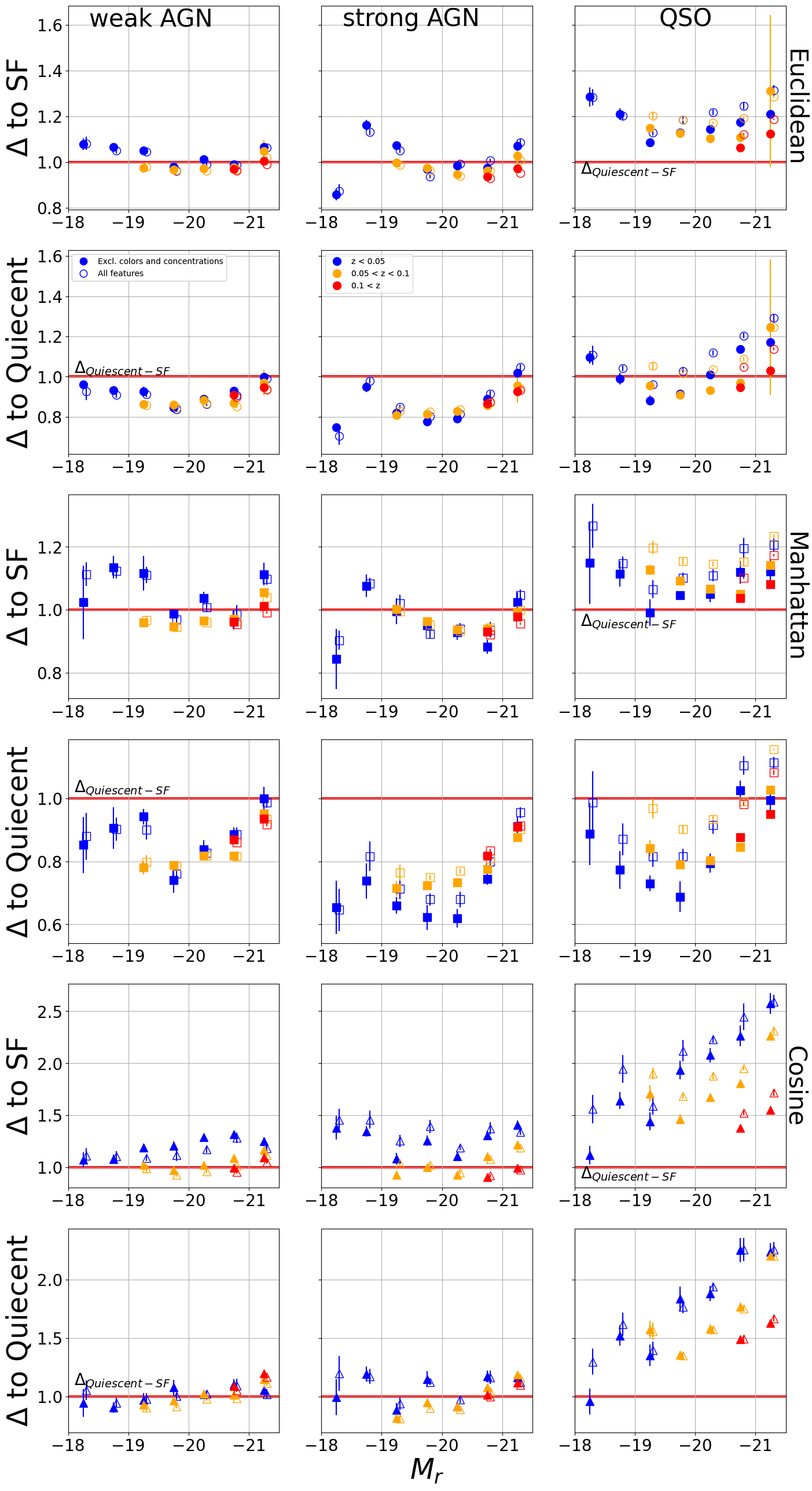}{0.5\textwidth}{}
          }
\caption{Distances from SF (first row) and Q (second row) types to weak\,AGN (left), strong\,AGN (center), and QSO (right) types in Euclidean (top two), Manhattan (middle two), and Cosine (bottom two) metrics. The color of a symbol represents the redshift range of a sample. The solid circles are distances on the morphological feature space of Table~\ref{tab:feat}. The open circles are distances measured by adding colors and concentrations, which are features sensitive to the AGN component.
\label{fig:dist_mag_separate}}
\end{figure}

QSO types have the lowest probability of corresponding to any of the morphological classes, which \citet[][Section~4.1]{Huertas-Company11} state means that the galaxy is not close to any of the classes of the training. The QSO type is identified with a QSO template, which differs from the $k$-corrected $g-r$ and $r-i$ colors of the galaxy spectra. Still, the QSO type has distances farther from the SF type than the Q type. Figure~\ref{fig:dist_mag_separate} shows the distances to the Q and SF types separately from both the AGN and QSO types normalized by the distances between the Q and SF types. The QSO type has distances to the Q type 90\%--130\% of $\Delta_{Q-SF}$ and the distances to SF types 105\%--130\% of $\Delta_{Q-SF}$.
The distances from QSO and to both Q and SF types increase by $\sim$10\% when we take into account the colors and concentration features (empty circles). In contrast, the distance changes from AGN types to Q and SF types are not significant.

\section{Discussion} \label{sec:dis}

The close proximity of AGN/QSO types to Q types in the morphological feature space is consistent with negative AGN FB suppressing star formation in host galaxies and turning galaxies to Q. However, it remains a theoretical challenge to explain how radiative pressure or outflow jets from the accretion disk can result in changes to the morphologies of host galaxies directly \citep{Vayner21, Couto23} although some studies have demonstrated that this may be possible \citep{Bi25, Goddard25}. Alternatively, a third mechanism could have played a role in the morphological transition and affected the ignition of AGN and/or the depletion of SF, such as a galaxy merger \citep[][but see \citealt{Gabor09}]{Weigel18} or a large-scale structure \citep{Silverman08} or both \citep{Erostegui25}. In the case of the large-scale structure, a distance-environment density relation herein marginally shows that the AGN types tend to have a closer distance to Q at a higher density (see Figure~\ref{fig:dist_dens}). On the other hand, \citet{Bichang24} could not relate dwarf galaxies' AGN activities to any other properties, such as local environments \citep[see also][]{Kaviraj25}, galaxy interactions, and a prompt quenching of the SF.

The distance ratios $\Delta$\,to\,SF\,/\,$\Delta$\,to\,Q from AGN types reaches its highest values at $-19$\,$>$\,$M_r$\,$>$\,$-20$ and decrease as the magnitude decreases. We matched the number of Q and SF types to the total number of AGN and QSO types in each sample bin so that the distance ratio values are not affected by the different numbers of Q and SF types. Assuming $\Delta$\,to\,SF\,/\,$\Delta$\,to\,Q is a proxy for the negative efficiency of AGN FB for each AGN object in each magnitude bin, we multiply the fraction of AGN derived from WISE colors to estimate the total amount of AGN FB: 
\begin{equation}
\mathrm{Effective\,AGN\,FB} = (\Delta\,to\,\mathrm{SF} / \Delta\,to\,\mathrm{Q}) \times f_{\mathrm{AGN}},
\end{equation}
where $\Delta$\,to\,SF\,/\,$\Delta$\,to\,Q values in the Euclidean metric are for weak and strong\,AGN from Figure~\ref{fig:dist_mag} and $f_{\mathrm{AGN}}$ from \citet[][thick green line in Figure~\ref{fig:frac}]{Kaviraj19}.
\begin{figure}
 \gridline{ 
            \fig{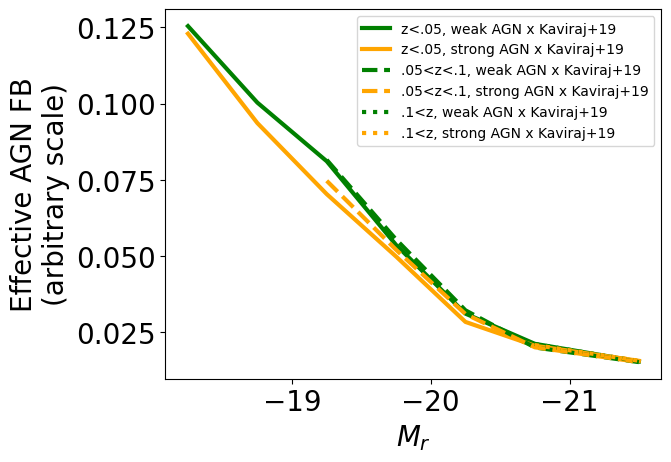}{0.5\textwidth}{}
          }
\caption{Effective FB amount that we estimate with the $\Delta$-to-SF/$\Delta$-to-Q Euclidean distance ratios as effectiveness multiplied by the fraction of each type in the mass bin. Both terms dominated in the dwarf regime, indicating the significance of the AGN feedback.
\label{fig:eff}}
\end{figure}
If the whole population hosting AGN is negligible, then the overall AGN FB would be small. Single-fiber spectroscopy studies, such as our sample, find $\sim$1\% of AGN or QSO types in dwarf galaxies, which could be biased by flux and fiber limitation. \citet{Penny18} found six among the 69 quenched low-mass galaxies ($\sim$9\% in $M_r$\,$>$\,$-19$) that appear to have an active AGN that prevents ongoing star formation using SDSS-\uppercase\expandafter{\romannumeral4} Mapping Nearby Galaxies at APO (MaNGA). \citet{Mezcua24} also used the MaNGA survey data and found the fraction of AGN $\sim$ 20\% in the dwarf population, which is similar to the result of \citet{Kaviraj19}. 

Figure~\ref{fig:eff} shows the effective AGN FB as a function of magnitude. The fraction of AGN calculated following \citet{Kaviraj19} is also highest in dwarf $\sim$\,14\% and falls to $\sim$\,2\% on the brightest end. Overall, the effective AGN FB estimate is more than an order of magnitude higher in the dwarf ($-18$\,$>$\,$M_r$\,$>$\,$-18.5$) than in the massive ($-21$\,$>$\,$M_r$) regime. Although the redshift range of the AGN fraction by \citet{Kaviraj19}, $z$\,$=$\,[0.1,\,0.3], is somewhat higher than that of the distance measurements, $z$\,$\lesssim$\,0.2, it is noteworthy that both the morphological feature distance and the AGN fraction suggest that AGN activity seems to have played a significant role in quenching SF not only in massive galaxies but also in dwarf galaxies.

\section{Summary}\label{sec:summary}

We measured multidimensional distances among Q-, SF-, AGN-, and QSO-type galaxies in the morphological feature space of axial ratios, model fit likelihoods, adaptive shape measures, and fourth moments. We find that
\begin{enumerate}
    \item Galaxies hosting AGN/QSO have morphological features more similar to Q than SF types.
    \item AGN/QSO and Q types are closer in feature space than SF types at lower host galaxy luminosities.
    \item The QSO types are clustered in the corners of the Q types on the \code{t-SNE} map. This suggests that the morphological features of the QSO type are similar to each other, and the features are closer to the Q type than other types.
    \item We obtain qualitatively similar results when considering Hubble types rather than the distances of morphological features. Using the continuous probabilities from \citet{Huertas-Company11}, we find that lower-luminosity galaxies with AGN are more likely to have the same Hubble types that Q galaxies have.
    \item Q-like morphological features of AGN types combined with their larger fractions in lower-mass galaxies suggest a significant role of the AGN FB in the dwarf regime.
\end{enumerate}

%% Please use the acknowledgment and contribution environments. This will 
%% be anonomyized when the "anonymous" style option is used. 
\begin{acknowledgments}
We thank the anonymous reviewer for the helpful comments that improved the manuscript.
D.K. acknowledges support from the National Research Foundation of Korea (NRF) grant funded by the Korean government (MSIT) (No. NRF-2022R1C1C2004506). 
G.~M. acknowledges support from the UK STFC under grant ST/X000982/1.
Funding for the Sloan Digital Sky Survey V has been provided by the Alfred P. Sloan Foundation, the Heising-Simons Foundation, the National Science Foundation, and the Participating Institutions. SDSS acknowledges support and resources from the Center for High-Performance Computing at the University of Utah. SDSS telescopes are located at Apache Point Observatory, funded by the Astrophysical Research Consortium and operated by New Mexico State University, and at Las Campanas Observatory, operated by the Carnegie Institution for Science. The SDSS website is \url{www.sdss.org}.
SDSS is managed by the Astrophysical Research Consortium for the Participating Institutions of the SDSS Collaboration, including Caltech, The Carnegie Institution for Science, Chilean National Time Allocation Committee (CNTAC) ratified researchers, The Flatiron Institute, the Gotham Participation Group, Harvard University, Heidelberg University, The Johns Hopkins University, L’Ecole polytechnique f\'{e}d\'{e}rale de Lausanne (EPFL), Leibniz-Institut f{\"u}r Astrophysik Potsdam (AIP), Max-Planck-Institut f{\"u}r Astronomie (MPIA Heidelberg), Max-Planck-Institut f{\"u}r Extraterrestrische Physik (MPE), Nanjing University, National Astronomical Observatories of China (NAOC), New Mexico State University, The Ohio State University, Pennsylvania State University, Smithsonian Astrophysical Observatory, Space Telescope Science Institute (STScI), the Stellar Astrophysics Participation Group, Universidad Nacional Aut\'{o}noma de M\'{e}xico, University of Arizona, University of Colorado Boulder, University of Illinois at Urbana-Champaign, University of Toronto, University of Utah, University of Virginia, Yale University, and Yunnan University.
\end{acknowledgments}

\begin{contribution}
%%This section gives authors the space to recognize author contributions. The text inside this environment is NOT counted towards the total word quanta. At a minimum, manuscripts are expected to include this text:

%All authors contributed equally to the Terra Mater collaboration.

%% But authors are expected to provide more specific details, e.g. 
%%
%%SC was responsible for writing and submitting the manuscript.
D.K. came up with the initial research concept and provided the formal analysis.
G.M. provided the validation and edited the manuscript.
%%OTS obtained the funding and edited the manuscript.
%%EBF provided the formal analysis and validation. He also edited the manuscript.
%%GEH Supervised the undergraduates, wrote the software and administers the project github and Zenodo repositories.
%%
%% Authors can use the Contributor Role Taxonomy (CRediT) at
%% https://credit.niso.org
%% for ideas on how write a good statement tailored to their needs.

\end{contribution}

%% To help institutions obtain information on the effectiveness of their 
%% telescopes the AAS Journals has created a group of keywords for telescope 
%% facilities.
%
%% Following the acknowledgments section, use the following syntax and the
%% \facility{} or \facilities{} macros to list the keywords of facilities used 
%% in the research for the paper.  Each keyword is check against the master 
%% list during copy editing.  Individual instruments can be provided in 
%% parentheses, after the keyword, but they are not verified.
\facilities{Sloan.}

%% Similar to \facility{}, there is the optional \software command to allow 
%% authors a place to specify which programs were used during the creation of 
%% the manuscript. Authors should list each code and include either a
%% citation or url to the code inside ()s when available.

\software{\code{astropy} \citep{astropy},
        \code{Orange} \footref{orange}.
          }

%% Appendix material should be preceded with a single \appendix command.
%% There should be a \section command for each appendix. Mark appendix
%% subsections with the same markup you use in the main body of the paper.
%%
%% Each Appendix (indicated with \section) will be lettered A, B, C, etc.
%% The equation counter will reset when it encounters the \appendix
%% command and will number appendix equations (A1), (A2), etc. The
%% Figure and Table counter will not reset.

%\appendix

%\section{Appendix information}

%% For this sample we use BibTeX plus aasjournalv7.bst to generate the
%% the bibliography. The sample7.bib file was populated from ADS. To
%% get the citations to show in the compiled file do the following:
%%
%% pdflatex sample7.tex
%% bibtext sample7
%% pdflatex sample7.tex
%% pdflatex sample7.tex

\bibliography{ms}{}
\bibliographystyle{aasjournalv7}

%% This command is needed to show the entire author+affiliation list when
%% the collaboration and author truncation commands are used.  It has to
%% go at the end of the manuscript.
%\allauthors

%% Include this line if you are using the \added, \replaced, \deleted
%% commands to see a summary list of all changes at the end of the article.
%\listofchanges

\end{document}